\documentclass[preprints,article,accept,moreauthors,pdftex]{Definitions/mdpi}

\pdfoutput=1

\usepackage{tikz}
\usepackage{booktabs}
\usepackage{siunitx}

\firstpage{1} 
\pubvolume{1}
\issuenum{1}
\articlenumber{0}
\pubyear{2022}
\copyrightyear{2022}
\datereceived{} 
\dateaccepted{} 
\datepublished{} 
\hreflink{https://doi.org/} 

\Title{Abiogenesis on Different Star Types; a Dissipative Photochemical Perspective}

\TitleCitation{Abiogenesis on Different Star Types}

\Author{Andrés Ledesma$^{1}$  and Karo Michaelian$^{2,\orcidA{}}$}

\AuthorNames{Andrés Ledesma, Karo Michaelian}

\AuthorCitation{Ledesma, A.; Michaelian, K.}

\address{%
\quad [1] Faculty of Science, Universidad Nacional Aut\'onoma de M\'exico, Circuito Interior de la Investigaci\'{o}n Cient\'{i}fica, Ciudad Universitaria, Ciudad de M\'{e}xico, C.P. 04510.; andreslr@ciencias.unam.mx\\
\quad [2] Department of Nuclear Physics and Application of Radiation, Instituto de F\'{i}sica, Universidad Nacional Aut\'onoma de M\'exico, Circuito Interior de la Investigaci\'{o}n Cient\'{i}fica, Ciudad Universitaria, Ciudad de M\'{e}xico, C.P. 04510.; karo@fisica.unam.mx}

\corres{karo@fisica.unam.mx}

\abstract{From the non-equilibrium thermodynamic perspective of the origin of life as a photochemical dissipative structuring (entropy driven) process, we assess the probability of carbon-based life arising on Earth-like analogues orbiting different main-sequence stellar types (O7 V to M2 V). Using black-body spectra normalized to Earth’s solar constant, we calculate surface photon fluxes for an atmosphere like early Archean Earth's in the productive dissipative structuring ($P$, soft UV-C + UV-B, 205–320 nm) and destructive ionization ($D$, hard UV-C + EUV, $<205$ nm) regions. Stationary concentrations of fundamental molecules and times to reach 99\% of these are computed for different chemical degradation (e.g., deamination,  hydrolysis, oxidation, etc.) rate constants of, $k = 10^{-7}$, $10^{-6}$, and $10^{-5}$ s$^{-1}$. For a nominal chemical degradation rate constant of $k = 10^{-6}$ s$^{-1}$ ($t_{1/2}= 8$ days), results show F-, G-, and K-type stars provide the highest stationary concentrations of fundamental molecules and short rise times (weeks to months), while quiescent M-type stars yield extremely low concentrations ($\sim10^{-7}$ relative to G stars) and require years to reach even these values. Flaring M stars improve stationary concentrations by about an order of magnitude ($\sim10^{-6}$ relative to G stars) but produce adverse planet surface environments for complex evolution through dissipative structuring. From this non-equilibrium thermodynamic perspective, carbon-based life like Earth's is to be found most probably on F-, G-, and high-mass K-type stars, with intelligent life arising only on G-type stars. Low mass K- and M-dwarfs are highly unlikely to harbor life unless seeded via panspermia. Biosignatures related to the thermodynamic imperative of photon dissipation are proposed.}



\begin{document}

\section{Introduction}

Our Sun, a G-type star, represents fewer than 7\% of stars in the Milky Way. A central question in astrobiology, therefore, is whether carbon-based life, as we know it on Earth, could also arise on planets around stars of other spectral types. Answering this question requires understanding the specific physicochemical processes that led to the emergence of life on Earth sometime between 3.85 and 4.2 billion years ago \cite{MojzsisEtAl1996, BellEtAl2015, MoodyEtAl2024}. 

A growing amount of evidence indicates that the origin of life on Earth may have depended critically on UV photochemical synthesis processes \cite{Ferris1966, FerrisOrgel1966, SaganKhare1971, BossardEtAl1982, PownerEtAl2009, BarksEtAl2010, Michaelian2011, Ruiz-BermejoEtAl2013, PatelEtAl2015, MeinertEtAl2016, AirapetianEtAl2020, GreenEtAl2021, RimmerEtAl2021, KufnerEtAl2023}. Other studies emphasize the possible role of ultraviolet selection in favoring prebiotically relevant molecules \cite{Sagan1973, MulkidjanianEtAl2003, PownerEtAl2009, KufnerEtAl2023}, while still other studies evaluate star spectral types in supporting particular photochemical synthesis routes \cite{RanjanSasselov2016, RimmerEtAl2018}.

However, the origin of life, being a non-equilibrium thermodynamic process, was much more than just the synthesis (photochemical or otherwise) of biologically relevant molecules. Analogously, the origin of a hurricane is much more than just the sunlight-induced evaporation of ocean surface water. Life and hurricanes are examples of complex dissipative structures arising ``spontaneously'' and evolving to dissipate an external generalized thermodynamic potential; the surface solar photon spectrum, and an ocean surface - upper atmosphere temperature gradient, respectively. The ``spontaneous'' structuring of material under such generalized potentials, occurring as a thermodynamic response to increase the rate of dissipation of that potential, is a well-understood non-equilibrium thermodynamic phenomenon in the non-linear region \cite{GlansdorffPrigogine1971}. 

For non-linear relations between applied forces and corresponding flows in systems far-from-equilibrium, many possible, highly organized, stationary states (structuring) may be available to the system, each with a different rate of dissipation. The the initial or boundary conditions do not uniquely specify the non-equilibrium state to which the system will evolve, this is determined stochastically by fluctuations near critical points, but the general tendency for autocatalytic (positive feedback) non-linearity is towards states of greater rates of dissipation of the imposed potential (greater entropy production) \cite{Prigogine1967, GlansdorffPrigogine1971, NicolisPrigogine1977, KondepudiPrigogine1998}.

The ``Thermodynamic Dissipation Theory of the Origin of Life'' (TDTOL) \cite{Michaelian2009, Michaelian2011, Michaelian2021, Michaelian2023, Michaelian2026b}, asserts that abiogenesis on Earth was the emergence of such a non-equilibrium thermodynamic dissipative structuring process \cite{Michaelian2017} arising ``spontaneously'' to dissipate the soft UV-C (205-285 nm) surface photon flux of the early Archean. From simple, ubiquitous, and weakly-absorbing precursor molecules such as hydrogen cyanide (HCN), cyanogen (NCCN), and carbon dioxide (CO$_2$) dissolved in water, dissipative structuring under UV-C light resulted in a set of complex strongly absorbing and rapidly dissipating UV-C chromophores (nucleobases, aromatic amino acids, cofactors, co-enzymes and key pigments - the fundamental molecules of life). These, in turn, through mutual interaction and association (e.g., excitation energy and charge transfer, auto-catalysis and surface anchoring), under the same UV-C light potential, evolved into still more complex systems increasing further the dissipation of the soft UV-C and soft UV-B (310-320 nm) Archean surface spectrum. This thermodynamic imperative thus provides an explanation for abiogenesis and the evolution of life grounded on physical-chemical law \cite{Michaelian2023}.

Photon wavelengths shorter than 205\,nm could ionize, dissociate, or otherwise degrade organic molecules, while wavelengths longer than 285\,nm lack sufficient energy to excite and reconfigure most double covalent carbon bonds. Wavelengths exceeding 320\,nm cannot directly drive the reconfiguration of single covalent bonds required for minimal dissipative structuring. 

Abiogenesis and subsequent evolution were thus entropy driven processes, leading to  fundamental organic molecules having; 1) high Gibb's free energies and large concentrations (compared to equilibrium values) \cite{Michaelian2021, MichaelianRodriguez2019}, 2) large molar extinction coefficients in the soft UV-C region (Figure \ref{fig:Pigments}), 3) large broadband wavelength absorption (Figure \ref{fig:DNA-Abs}), 4) high quantum yields to conical intersections which dissipate the electronic excitation energy extremely rapidly into heat (Figure \ref{fig:ConicalInt}), and 5) inter-molecular associations fomenting dissipation and leading to polymerization, enzymeless replication \cite{MichaelianSantillan2019}, homochirality \cite{Michaelian2018}, codon-amino acid relationships \cite{MejiaMichaelian2020}, etc., leading eventually to multicellular organisms and ecosystems \cite{Michaelian2005}, and 6) the ability to couple to abiotic dissipative processes such as the water cycle \cite{Michaelian2012b}, all driven by the non-equilibrium thermodynamic imperative of increasing the global photon dissipation rate. The result is today’s highly efficient planetary-scale dissipative system hosting photon dissipative capacity to well beyond the red edge ($\sim$700\,nm); the biosphere.

In contradistinction to the simple photochemical feedstock scenario, the TDTOL provides a complete framework for the processes of the origin and evolution of life, leading to the above mentioned extraordinary optical properties of the fundamental molecules, their co-evolution into associations, and coupling to abiotic dissipative processes, promoting still greater photon dissipation. It will be shown here that the origin and maintenance of these processes depends critically on a planet surface stellar light environment providing intense and continuous soft UV-C and diminished hard UV-C light.

Under the TDTOL, UV-C light was thus not merely an available energy source for synthesizing prebiotic molecules (the photochemical feedstock scenario), but rather it was the applied thermodynamic potential driving the entire dissipative structuring process of abiogenesis and biological evolution. If TDTOL correctly describes abiogenesis and evolution on Earth—as supported by evidence to be presented in Section~\ref{sec:Evidence}—then continuous UV light in a well defined spectral window must reach a planetary surface: bounded at short wavelengths by aqueous molecular ionization energies ($\lesssim$205\,nm) and at long wavelengths by the energies of carbon-based covalent bonds ($\lesssim$320\,nm). Planets orbiting stars that fail to deliver a significant and continuous flux in this spectral region, or an abundance of light within the degradation region ($< 205$ nm), are therefore unlikely to initiate or sustain a pathway to an origin-of-life analogous to Earth’s.

The primary goal of the present study is, therefore, to identify which stellar spectral types can deliver a continuous and intense soft UV-C + UV-B (205--320\,nm) surface flux while limiting light in the degradation region ($< 205$ nm), and therefore be promising hosts for an abiogenesis similar to Earth's. We restrict our analysis to Archean Earth-like planets located at the distance from their star that yields a top-of-atmosphere bolometric flux equal to the solar constant (1366\,W\,m$^{-2}$). This normalization ensures the presence of liquid water and broadly comparable photochemical and thermal reaction rates across the star types. Temperature dependent chemical degradation of the fundamental molecules (e.g., hydrolysis, deamination, oxidation, etc.) is also taken into account.

This analysis is especially timely since current and upcoming space telescopes (JWST, Habitable Worlds Observatory, LIFE, etc.) are actively characterizing exoplanet atmospheres in search of biosignatures. Observing time on these flagship facilities is severely limited, making it essential to prioritize targets with the highest a priori probability of hosting Earth-like life—and to know which specific thermodynamic and photochemical biosignatures to search for. 

\section{The Atmospheric UV-C Window}
\label{sec:EarlyEarth}

Although little empirical data regarding rocky exoplanet atmospheres exists (JWST could change this shortly), calculations assuming phase equilibrium for a given pressure, temperature, and element abundances (continental crust, bulk silicate Earth, mid oceanic ridge basalt, CI chondrites and abundances measured in polluted white dwarfs) suggest that, for a secondary out-gassed atmosphere and surface temperatures between $\sim 600$ and $\sim 3500$ K, the near-crust atmospheres of rocky planets would mainly be composed of H$_2$O, CO$_2$, and SO$_2$ and in some cases of O$_2$ and H$_2$, while for temperatures $\le 500$ K, N$_2$-rich or CH$_4$-rich atmospheres are predicted with trace amounts of H$_2$O, CO$_2$, and SO$_2$ \cite{HerbortEtAl2020}.

Atmospheric modeling suggests that planets on pre-main sequence low-mass stars (K/M dwarfs) may retain dense envelopes of primordial H$_2$ and He, persisting over Gyr timescales and resulting in a variety of secondary atmospheres, greenhouse effects, and habitability prospects \cite{OwenMohanty2016}. Water loss during intense early XUV radiation on planets at the pre-main-sequence phase of young M dwarfs can lead to thick secondary, Venus-like, CO$_2$-dominated atmospheres through out-gassing \cite{CaroneEtAl2025}, or atmospheres having an important O$_2$ component \cite{LugerBarnes2015}. Such atmospheres could shift the photon transmission window, but our contention is that any atmosphere shielding the surface from soft UV-C light would not activate the non-equilibrium thermodynamic mechanisms for abiogenesis and evolution which ostensibly occurred on Earth \cite{Michaelian2018, MichaelianSantillan2019, MejiaMichaelian2020, Michaelian2021}.

Early Archean Earth's atmospheric O$_2$ levels were less than $10^{-6}$ times present values, while N$_2$ levels were similar to today´s or possibly a few times lower. CO$_2$ and CH$_4$ levels were between $\sim$10 to 2500 and 10$^2$ to 10$^4$ times modern amounts, respectively \cite{Kasting1993, LoweTice2004, CatlingZahnle2020}. Greenhouse gas concentrations were sufficient to offset an approximately 25\% fainter Sun. A faster rotating Earth ($\sim$ 14 hour days) meant a greater latitudinal gradient in surface temperature \cite{SpaldingFischer2019} which may have ranged from $\sim 90^\circ$C at the equator \cite{SchwartzChang2002, Knauth1992, KnauthLowe2003} to 0$^\circ$C at the poles \cite{CatlingZahnle2020}.

Few photons of wavelength $< 205$ nm and between 285 and 305 nm reached Earth's surface during the Archean due to absorption by atmospheric  CO$_2$, N$_2$ and the aldehydes, respectively \cite{Sagan1973}. This photon potential was available at Earth's surface before the origin of life ($\sim$ 3.9 - 4.2 Ga) and persisted for at least 1200 million years thereafter, until the emergence of oxygenic photosynthesis and an ensuing UV ``protective'' ozone layer at around 2.7 Ga \cite{Sagan1973, CnossenEtAl2007}. 

For our analysis of rocky planets in the habitable zones, trace amounts (e.g., $<100$ ppm) of H$_2$O, CO$_2$, CH$_4$, H$_2$S and SO$_2$ define well the hard UV-C cut-off at about $\sim 205$ nm for atmosphere pressures comparable to that of Earth's Archean atmosphere \cite{Sagan1973} (similar to, or up to twice, present pressure, as determined from Archean rain drop imprints \cite{SomEtAl2012}). For these planets, the lack of reactive oxygen (and, therefore, also ozone) would lead to an atmospheric window for the soft UV-C surface flux required for dissipative structuring and a significantly diminished hard UV-C flux.

\section{The Precursors}
\label{sec:Precursors}

Hydrogen cyanide (HCN) was recognized over a century ago as a likely precursor of life's fundamental molecules \cite{Pfluger1875}, in particular of the nucleobases \cite{FerrisOrgel1966, Michaelian2021}. The formation of HCN in an N$_2$-rich atmosphere requires first breaking the triple covalent bond between nitrogens, N$\equiv$N, and then atomic nitrogen attacking a carbon atom. Given the atmospheric abundance of N$_2$ and a carbon to oxygen ratio of C/O$\ge$1 of the early Archean, this can be readily accomplished via photochemistry \cite{RimmerRugheimer2019}. $\mathrm{N}_2 $ absorbs strongly between approximately 120 to 145 nm. The $ \mathrm{N}_2 $ photodissociation energy $\sim$9.8 eV corresponds to a wavelength of 126.5 nm (close to the solar Lyman-$\alpha$ line of 121.6 nm). 

The removal of a proton from HCN in neutral water requires a photon of about 238.4 nm (5.2 eV) while its gas-phase photodissociation energy is about 13.3 eV (93 nm). HCN absorbs most strongly at $\sim$150 nm (8.27 eV). It has been estimated that HCN concentrations as high as $6\times 10^{-5}$ M may have been common in the enriched microlayer of the Archean ocean surface  \cite{Michaelian2021}.

\section{Evidence for a Photochemical Dissipative Origin of Life}
\label{sec:Evidence}

\begin{figure}[ht!]
\includegraphics[width=17cm]{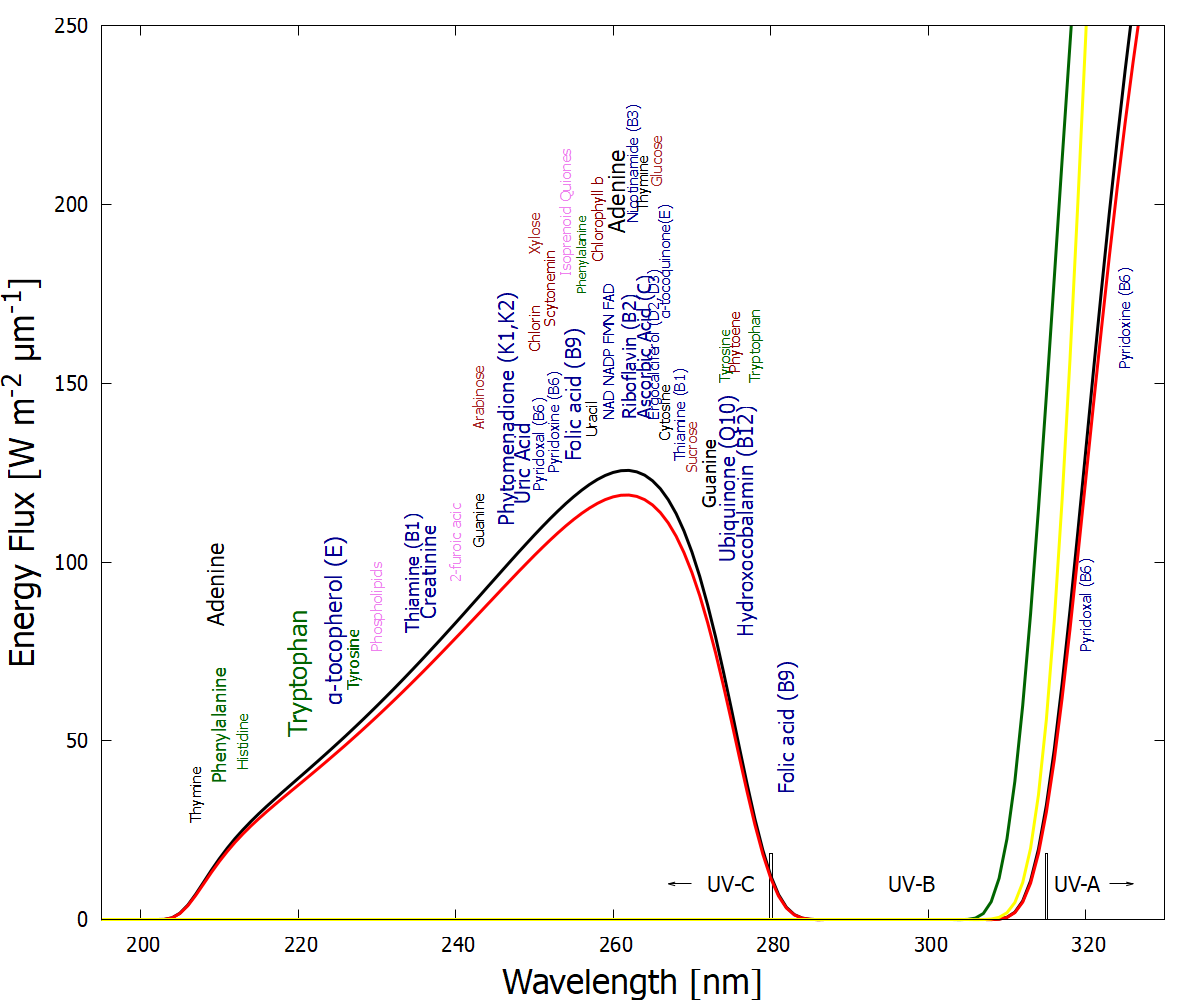}
\caption{The spectrum of UV light available at Earth's surface before the origin of life at approximately 3.9 Ga and until at least 2.9 Ga (curves black and red, respectively). This spectrum in the UV-C may even have persisted throughout the entire Archean until 2.5 Ga \cite{MeixnerovaEtAl2021}. Atmospheric CO$_2$, H$_2$O, SO$_2$ and probably some H$_2$S, were responsible for the absorption of wavelengths shorter than $\sim$205 nm, and atmospheric aldehydes (e.g., formaldehyde and acetaldehyde, common photochemical products of CO$_2$ and water) absorbed between about 285 and 305 nm \cite{Sagan1973, MellerMoortgat2000}), approximately corresponding to the UV-B region (280 and 315 nm). By around 2.2 Ga (green curve), UV-C light at Earth's surface was completely extinguished by the pigments of oxygen and ozone resulting from organisms performing oxygenic photosynthesis. The yellow curve corresponds to the present surface spectrum. \mbox{Energy fluxes} are for the Sun at the zenith. Over 50 of the fundamental molecules of life are plotted at their wavelengths of maximum absorption: nucleic acids (black), amino acids (green), fatty acids (violet), sugars (brown), vitamins, co-enzymes, and cofactors (blue), and pigments (red). We have suggested that these molecules were dissipatively structured as UV-C pigments under this light. The font size is roughly proportional to the relative size of the respective molar extinction coefficient of the pigment. Adapted with permission from Michaelian and Simeonov \cite{MichaelianSimeonov2015}.}
\label{fig:Pigments}
\end{figure}

\begin{figure}[ht!]
\includegraphics[width=10 cm]{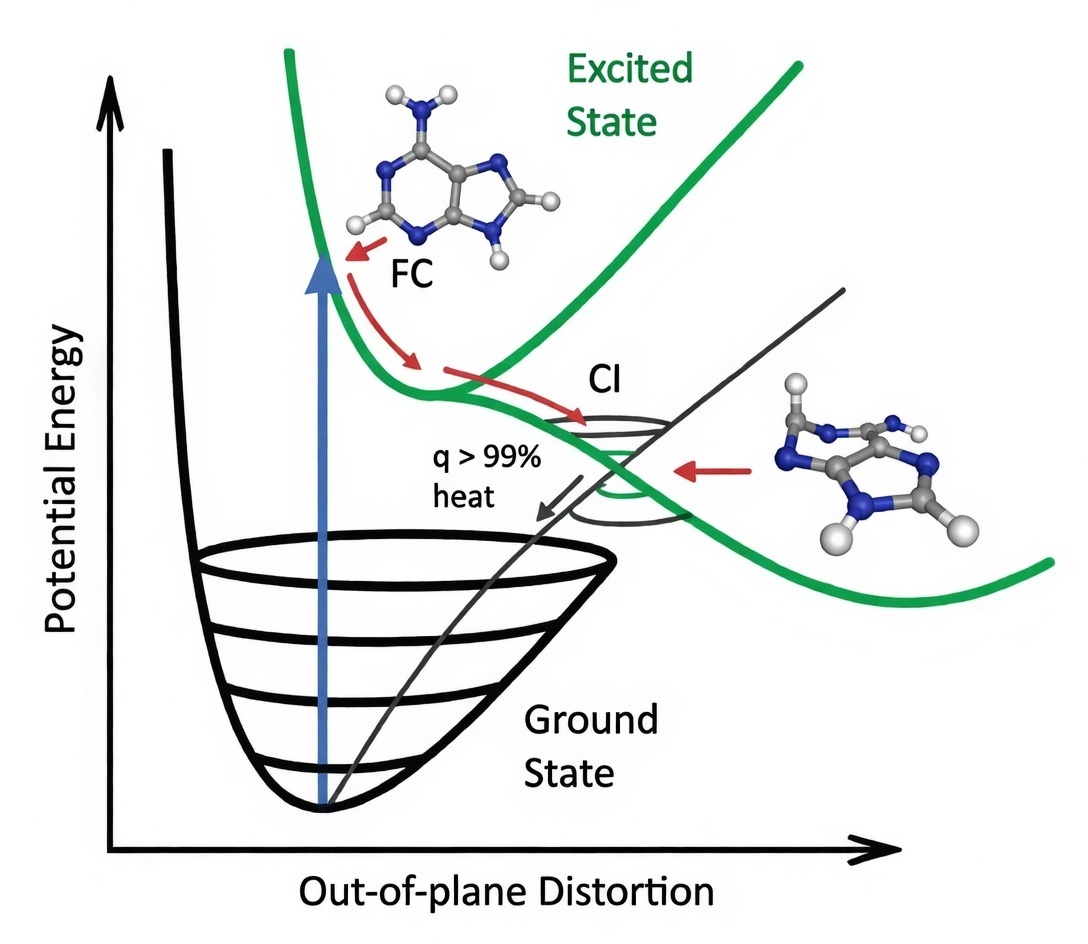}
\caption{A conical intersection (CI) for excited adenine showing a degeneracy of the electronic excited state with the vibrational states superimposed on the electronic ground state after a UV-C photon absorption event (blue arrow) which induces a nuclear coordinate deformation from the molecule's original planar structure in the Franck-Condon (FC) region to activation of an N9--H stretch or a ring-puckering motion known as {\em pyramidalization} (shown in the diagram). The most probable deformation depends on the incident photon energy and protonation state. It is this deformation, resulting from excitation of an anti-bonding state (e.g., $\pi \rightarrow \pi^*$), which leads to a lowering of the excited potential energy surface such that it intersects with vibrational states of the electronic ground state, resulting in the conical intersection. Conical intersections provide rapid (sub-picosecond) dissipation of the electronic excitation energy into vibrational energy (heat). The~quantum efficiency, $q$, for this dissipative route is very large ($> 99$\%) for many of the fundamental molecules of life, making them photochemically stable and, more importantly (from our thermodynamic perspective), very efficient at UV-C photon dissipation. Another common form of coordinate transformation mediated through conical intersections are proton and electron transfers within the molecule or with the solvent environment and this may have relevance to enzyme-less photon-induced denaturing of RNA and DNA \cite{MichaelianSantillan2019} and to photosynthesis. The diagram is based on data from Andrew Orr-Ewing~\cite{Orr-Ewing}, Roberts~et~al.~\cite{RobertsEtAl2014}, Kleinermanns~et~al.~\cite{KleinermannsEtAl2013}, and Barbatti~et~al.~\cite{BarbattiEtAl2010}). Reprinted with permission from Michaelian \cite{Michaelian2021}.}
\label{fig:ConicalInt}
\end{figure}

\begin{figure}[ht!]
\centering
\includegraphics[width=8 cm]{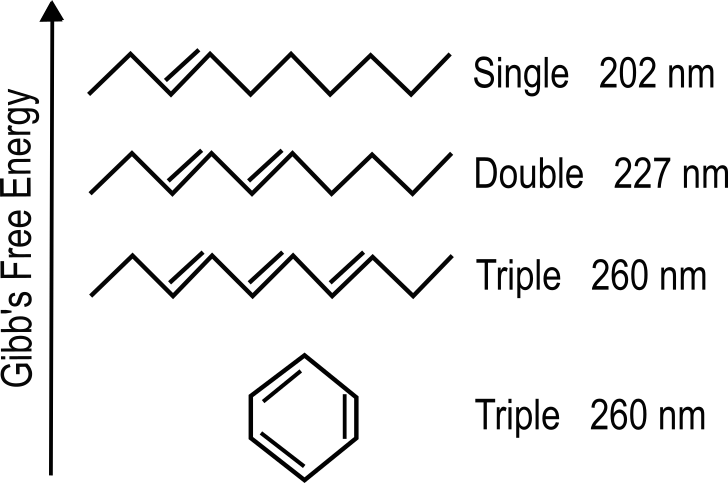} 
\caption{Conjugated carbon molecules are more stable (lower Gibb's free energy in the ground state) as compared to saturated molecules, but more importantly provide new collective electron orbitals giving rise to excited states at energies adequate for the absorption of soft UV-C photons. The greater the conjugation number, the greater the wavelength of maximum absorption. The wavelength of maximum absorption of the chromophore can, therefore, be tuned simply by a  protonation or deprotonation event. Conjugation is also important for giving molecules a conical intersection (Fig. \ref{fig:ConicalInt}) allowing rapid dissipation of the electronic excited state energy into heat (internal conversion). Under TDTOL, photon dissipation is the thermodynamic reason for abiogenesis. Reprinted with permission from Michaelian \cite{Michaelian2017}}.
\label{fig:Conjugation}
\end{figure}

\begin{figure}[ht!]
\centering
\includegraphics[width=15 cm]{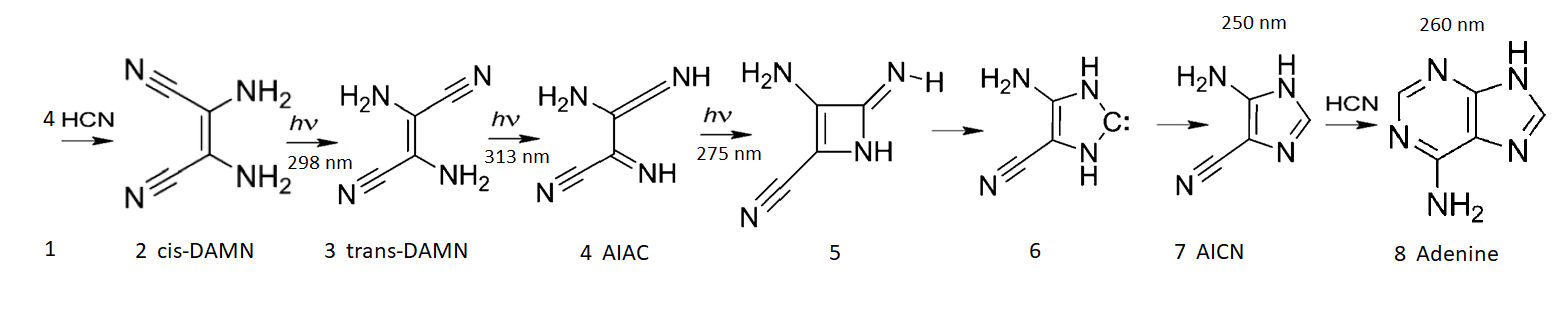} 
\caption{The photochemical dissipative structuring of adenine from 5 molecules of hydrogen cyanide (HCN) in water, first observed by Ferris and Orgel (1966) \cite{FerrisOrgel1966,BoulangerEtAl2013}. Four molecules of HCN (1) are transformed into the smallest stable oligomer (tetramer) of HCN, known as  cis-2,3-diaminomaleonitrile (cis-DAMN) (2), which, under a constant UV-C photon flux, isomerizes into trans-DAMN (3) (also known as diaminofumaronitrile, DAFN) which can be converted, on absorbing two more UV-C photons, into an imidazole intermediate, 4-amino-1H-imidazole-5-carbonitrile (AICN) (7).  Hot ground state thermal reactions with another HCN molecule or its hydrolysis product formamide (or ammonium formate) leads to the purine adenine (8). This is a microscopic dissipative structuring process which ends in adenine \cite{Michaelian2017, Michaelian2021}, a UV-C pigment with a large molar extinction coefficient at the maximum intensity of the UV-C Archean surface solar spectrum (260 nm - Figs. \ref{fig:Pigments}) and a peaked conical intersection facilitating rapid dissipation of photons at these wavelengths. The other nucleobases have similar optical characteristics and also appear to be UV-C molecular dissipative structures (e.g., reference \cite{HernandezMichaelian2022} and Figure \ref{fig:DNA-Abs}). Reprinted with permission from Michaelian \cite{Michaelian2021}.}
\label{fig:AdenineSyn}
\end{figure}

\begin{figure}[ht!]
\includegraphics[width=13cm]{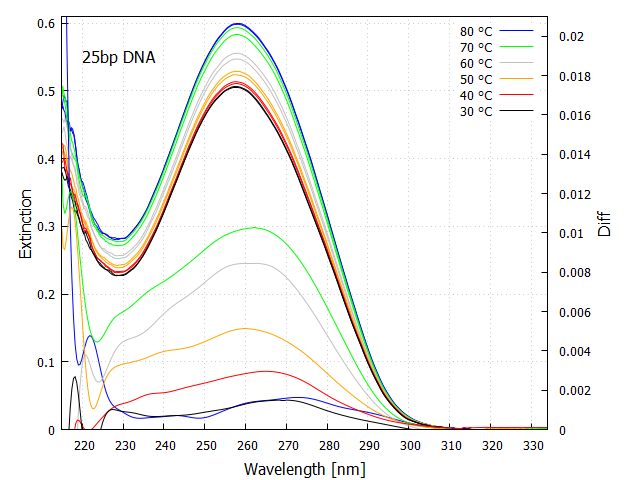}
\caption{Absorption spectrum of 25bp DNA (including all DNA nucleobases) in the soft UV-C region showing hyperchromism resulting from the denaturing with temperature. The lower curves are the thermal difference spectra (right y-axis) obtained by subtracting the lower temperature extinction curve from the higher temperature curve for a 3 $^\circ$C bin centered at the specified temperature (smoothed with a 2000 point Bezier function). Peak absorption at $\sim$260 nm corresponds to the peak in the incident UV-C spectrum arriving at Earth's surface during the Archean (Fig. \ref{fig:Pigments}). Large broadband absorption implies rapid (sub-picosecond) dissipation of the electronic excitation energy into heat through conical intersections (Fig. \ref{fig:ConicalInt}). Reprinted with permission from Michaelian and Santillan \cite{MichaelianSantillan2019}. }
\label{fig:DNA-Abs}
\end{figure}

Empirical evidence supports our assertion that the fundamental molecules of life were initially dissipatively structured UV-C chromophores (the ``Pigment World'' hypothesis  \cite{Michaelian2011, Michaelian2024}) and that the origin and evolution of life was a non-equilibrium process driven by  dissipating continuous sunlight:
\begin{enumerate}
\item Non-equilibrium processes are driven by a continuous source of free energy. The free energy available in UV light of wavelength less than 320 nm arriving at Earth's surface today is more than 1000 times greater than that of all other non-photon energy sources combined \cite{MillerOrgel1974}, and this would have been even greater during the Archean because of the lack of an atmospheric ozone layer.

\item The wavelength of maximum absorption of many of the fundamental molecules coincide with the predicted UV-C window in the Archean atmosphere (Fig. \ref{fig:Pigments}) and this coincides neatly with the spectral region ($\sim$205--320 nm) required for dissipative structuring.

\item Many of the fundamental molecules of life are endowed with {\em peaked conical intersections} giving them broad band absorption and large quantum yields for internal conversion, i.e. extremely rapid (picosecond) dissipation of the photon-induced electronic excitation energy into vibrational energy of the atomic coordinates, and finally into the surrounding water solvent \cite{Michaelian2017, SchuurmanStolow2018, Michaelian2021}. The conical intersection is created by a photon-induced excitation of an electron into an anti-bonding orbital (e.g., $\pi \rightarrow \pi^*$) which weakens the respective bond, decreasing the energy of the excited state upon elongation of the bond, leading to intersection of the excited state potential energy surface with that of the electronic ground state (Fig. \ref{fig:ConicalInt}).

\item Even minor transformations (e.g., protonations, tautomerizations, oxidation/reduction or methylations) of the fundamental molecules of life, which often endows them with lower Gibb's free energy, eliminates completely, or significantly reduces, their extraordinary photon absorption and dissipation properties \cite{CohenEtAl2003}.

\item The wavelength of maximum absorption of the fundamental molecules can be tuned simply by a protonation or deprotonation event (changing the conjugation number), decreasing or increasing, respectively, this wavelength by about 30 nm (Fig. \ref{fig:Conjugation}). This allows photon dissipative chromophores to easily ``evolve'' towards dissipation of the higher intensity light at longer wavelengths and thereby thermodynamically ``adapt'' towards dissipation of the most intense wavelengths, as well as ``track'' a possibly evolving surface spectrum.

\item Many photochemical routes, from common and simple Archean precursor molecules like hydrogen cyanide or cyanogen, to the synthesis of nucleic acids \cite{Ferris1966, FerrisOrgel1966}, amino acids \cite{SaganKhare1971}, fatty acids \cite{MichaelianRodriguez2019}, sugars \cite{Ruiz-BermejoEtAl2013, MeinertEtAl2016}, chlorophyll \cite{MichaelianSimeonov2026} and other pigments \cite{MichaelianSimeonov2015, SimeonovMichaelian2025} have been identified at these UV-C wavelengths.

\item The rate of photon dissipation within the Archean UV-C window generally increases after each incremental transformation on route to photochemical synthesis of the fundamental molecule (Fig. \ref{fig:AdenineSyn}), a hallmark of dissipative structuring in the non-equilibrium thermodynamic regime \cite{Michaelian2017, MichaelianRodriguez2019, Michaelian2021}.

\end{enumerate}

These extraordinary photochemical properties of the fundamental molecules of life are not arbitrary coincidences, but rather the expected attributes of molecular dissipative structuring in the soft UV-C spectral region.

\section{Photon-induced Molecular Degradation}

Sustained photon dissipative structuring of molecules is inhibited by photons of wavelengths shorter than approximately $205$ nm where ionization promoting molecular degradation may occur. The Earth's Archean surface spectrum, in fact, had relatively few photons of degradation ($D$) wavelengths shorter than about 205 nm compared to the dissipative production ($P$) wavelengths $205-320$ nm ($P/D\approx457$  - Fig.  \ref{fig:SpectraAtSurface} and Table \ref{tab:Photon_fluxes_Earth_surf}) since $D$ photons are strongly absorbed by trace amounts of atmospheric CO$_2$, H$_2$S, SO$_2$  and H$_2$O. Such trace gases are also likely to be found on rocky planets of other star types \cite{HerbortEtAl2020}. The ratio of $P/D$ at the surface is, however, strongly dependent on star type (Table \ref{tab:Photon_fluxes_Earth_surf}).

Ionization energies (IEs) listed in Table \ref{tab:IonizationE} of some of the important fundamental molecules in water at neutral pH were obtained from published gas phase values by applying a solvation correction which decreases IEs due to stabilization of the cation radical by the polar solvent. Studies suggest a 1.5–2 eV reduction for neutral bases \cite{CauetEtAl2007} and amino acids \cite{SchroederEtAl2015}, and 3–5 eV for nucleotides (with phosphates). For Table \ref{tab:IonizationE} we assumed a $\sim$1.9 eV shift for nucleobases (without phosphates) and amino acids, and $\sim$2.0 eV for sugars (interpolated from polyol studies \cite{KumariVaval2013}), reflecting typical small-molecule solvation \cite{BravayaEtAl2010}. At neutral pH, amino acids exist as zwitterions (except histidine’s imidazole, which is partially protonated) and nucleic bases are neutral.

\begin{table}[htb]
\centering
\caption{Estimated aqueous ionization energies (IEs) of nucleic acids, amino acids, sugars and fatty acids in water at pH 7, along with corresponding photon wavelengths ($\lambda$). The values listed are approximate, obtained from gas-phase experimental data adjusted by typical solvation shifts assuming neutral (pH 7) and physiological forms (zwitterions for amino acids and neutral bases). Charged states (e.g., deprotonated cysteine) could shift IEs further. Nucleobases are listed alone, not as nucleotides, and sugars are free, not in polymers. Including phosphates or glycosidic bonds would lower IEs further (e.g., AMP $\sim$4.5 eV). }
\begin{tabular}{lcc}
\toprule
Molecule & IEs (eV) & $\lambda$ (nm) \\
\midrule
\multicolumn{3}{c}{\textit{Nucleic Acid Bases}} \\
Adenine & 6.5 & 190 \\
Cytosine & 6.7 & 185 \\
Guanine & 6.0 & 207 \\
Thymine & 6.9 & 180 \\
Uracil & 7.1 & 175 \\
\midrule
\multicolumn{3}{c}{\textit{Selected Amino Acids}} \\
Alanine & 7.7 & 161 \\
Cysteine & 6.4 & 194 \\
Glycine & 7.6 & 163 \\
Histidine & 6.8 & 182 \\
Tryptophan & 7.3 &  170 \\
Tyrosine & 6.5 & 190 \\
\midrule
\multicolumn{3}{c}{\textit{Sugars}} \\
Ribose & 7.5 & 165 \\
Deoxyribose & 7.5 & 165 \\
Glucose & 7.7 & 161 \\
Fructose & 7.6 & 163 \\
\midrule
\multicolumn{3}{c}{\textit{Fatty Acids}} \\
\multirow{2}{*}{Short-Chain} & & \\
\ \\
Acetic Acid (C2:0) & 9.2 & 135 \\
Butyric Acid (C4:0) & 8.7 & 143 \\
\midrule
\multirow{2}{*}{Medium-Chain} & & \\
\ \\
Caprylic Acid (C8:0) & 8.3 & 149 \\
Capric Acid (C10:0) & 8.2 & 151 \\
\midrule
\multirow{2}{*}{Long-Chain} & & \\
\ \\
Palmitic Acid (C16:0) & 8.0 & 155 \\
Stearic Acid (C18:0) & 7.9 & 157 \\
$\alpha$-Eleostearic Acid (C18:3, 9c,11t,13t) & 7.4 & 168 \\
$\beta$-Eleostearic Acid (C18:3, 9t,11t,13t) & 7.4 & 169 \\
Punicic Acid (C18:3, 9c,11t,13c) & 7.4 & 166\\
\bottomrule
\bottomrule
\end{tabular}
\label{tab:IonizationE}
\end{table}

Short and medium-chain fatty acid IEs in water are derived from gas-phase data (acetic, butyric) and trends (caprylic, capric), reduced by $\sim$1.5 eV for aqueous anionic forms. Long-chain saturated fatty acid IEs are estimated from chain-length trends and reduced by $\sim$1.5 eV in water. The $\sim$1.5 eV reduction approximates deprotonation ($\sim$1 eV) and solvation ($\sim$0.5 eV), but actual shifts vary ($\pm$ 0.5 to 1 eV). For the triple conjugated fatty acids, the IEs are estimated $\sim$0.4–0.6 eV lower than stearic acid’s gas-phase value due to conjugation, then further reduced by $\sim$1.5 eV. Slight variations reflect cis/trans effects (all-trans $\beta$-eleostearic fatty acids are lowest).

All fundamental molecules listed in Table \ref{tab:IonizationE} have higher aqueous ionization energies than the long wavelength cutoff for degradation $D$ photons set to $\sim$205 nm. Guanine is near the edge at 207 nm.

\section{Star Type Spectral Outputs}

Table \ref{tab:Parameters} lists the stellar parameters of main-sequence O, B, A, F, G, K, M-type, luminosity class V, stars with planet-star distances normalized to the solar constant of 1366 W/m$^2$. The calculations use standard astronomical constants: $\sigma = 5.670367 \times 10^{-8}$ W/m$^2$/K$^4$, solar surface temperature 5772 K, and solar radius ($R_{\odot}$).

\begin{table}[ht!]
\centering
\caption{Stellar parameters and energy/photon fluxes at the top of an Earth-like atmosphere for a planet receiving the solar constant of \SI{1366}{\watt\per\meter\squared}.}
\label{tab:Parameters}
\small
\begin{tabular}{l c c c c c c c}
\toprule
Type & $T$ (K) & $R$ ($R_\odot$) & $d$ (au) & $\lambda_{\max}^E$ (nm) & $F_E$ (W/m$^2$/µm) & $\lambda_{\max}^P$ (nm) & $F_P$ (ph/m$^2$/s/µm) \\
\midrule
O7 V & 30000 & 6.60 & 178.3 & 96.6 & 7866 & 122.3 & $4.21\times 10^{21}$ \\
B2 V & 20000 & 4.19 & 50.31 & 144.9 & 5244 & 183.5 & $4.21\times 10^{21}$ \\
A2 V & 9000 & 1.86 & 4.515 & 322.0 & 2360 & 407.7 & $4.21\times 10^{21}$ \\
F5 V & 6500 & 1.39 & 1.763 & 445.8 & 1705 & 564.6 & $4.21\times 10^{21}$ \\
G2 V & 5772 & 1.00 & 1.000 & 502.0 & 1507 & 635.8 & $4.21\times 10^{21}$ \\
K2 V & 5000 & 0.673 & 0.5051 & 579.6 & 1312 & 733.9 & $4.21\times 10^{21}$ \\
M2 V & 3500 & 0.243 & 0.0892 & 827.9 & 918.7 & 1048.5 & $4.21\times 10^{21}$ \\
\bottomrule
\end{tabular}
\end{table}

Listed temperatures are typical effective temperatures for mid-range subtypes of main-sequence stars \cite{PecautMamajek2013}. Radii are typical for main-sequence stars and matched to the temperatures. Once on the main sequence, the emitted spectrum of a star remains relatively stable for most of its lifetime, depending basically on its mass and metallicity.

Planet-star distances $d$ (au) are normalized by integrating the energy flux and equating it to the solar constant (1366 W/m$^2$), or, for a black-body, more easily obtained from,
\[
\sigma T^4 \left( \frac{R}{d} \right)^2 = 1366
\]
giving,
\[
d_{\text{au}} = \left( \frac{R}{R_{\odot}} \right) \cdot \left( \frac{T}{5772} \right)^2.
\]

The electromagnetic spectrum emitted by bodies in thermal equilibrium is given by Planck's blackbody radiation equation and depends only on the temperature $T$. As a function of wavelength $\lambda$, the energy density (energy per unit volume per unit wavelength interval $d\lambda$) has the following form, 
\begin{equation} u(\lambda)d\lambda = \frac{8\pi hc }{\lambda^5 (e^{\beta hc / \lambda}- 1)}d\lambda,
\end{equation}
giving the amount of energy radiated between wavelengths $\lambda$ and $\lambda +d\lambda$ in a unit volume, with $\beta=1/kT$.

This energy distribution $u(\lambda)$ has a maximum as a function of temperature as determine by Wien's displacement law,

\begin{equation}
\lambda_{max, E} = \frac{b}{T},
\end{equation}
where $\lambda_{max, E}$ (nm) is the wavelength at which the peak of maximum energy emission occurs. $T$ is the absolute temperature of the object in units of K, and $b=2.8978 \times 10^6$ (nm$\cdot$K) is the proportionality constant known as Wien’s constant.

The spectral energy flux at $\lambda_{\text{max, E}}$ (W/m$^2$/$\mu$m) is computed using the Planck function, normalized to the planet distance,
\[
F(\lambda_{\text{nm}}) = \frac{3.131 \times 10^{23}}{\lambda_{\text{nm}}^5} \frac{1}{e^{1.438776877 \times 10^7 / (\lambda_{\text{nm}} T)} - 1} \frac{1366}{\sigma T^4}
\]
where $\lambda_{\text{nm}} = 2.8978 \times 10^6/T$. The constant $3.131 \times 10^{23}$ ensures $F(502)$ equals the solar output ($\approx 1507$ W/m$^2$/$\mu$m) for a G2 V star (Sun) at 502 nm.

\begin{figure}[ht!]
\includegraphics[width=13.8cm]{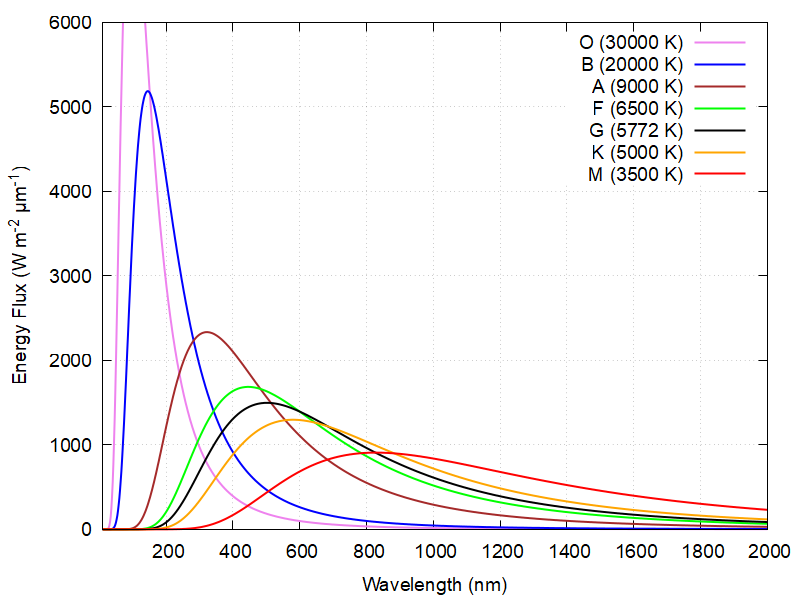}
\caption{The energy fluxes as a function of wavelength of different star types at the top of the atmosphere of a planet like Earth at a distance from its star that normalizes the wavelength integrated spectral constant to that of the Sun (1366 W/m$^2$).}
\label{fig:Spectral_Energy_Flux}
\end{figure}

In figure \ref{fig:Spectral_Energy_Flux}, the planet-star distance normalized spectral energy flux $F(\lambda_{\text{nm}})$ is plotted over the wavelength region 0--2000 nm for all star types.

The wavelength of maximum photon flux (nm) is determined as,
\[
\lambda_{\text{max, P}} = \frac{3.6697 \times 10^6}{T}
\]
derived from the photon number flux peak, adjusted for Wien's law.

The photon flux at $\lambda_{\text{max, P}}$ (photons/m$^2$/s/$\mu$m) is,
\[
\Phi(\lambda_{\text{nm}}) = \frac{1.576 \times 10^{39}}{\lambda_{\text{nm}}^4} \frac{1}{e^{1.438776877 \times 10^7 / (\lambda_{\text{nm}} T)} - 1} \frac{1366}{\sigma T^4}
\]
At $\lambda_{\text{nm}} = 3.6697 \times 10^6/T$, $\Phi_{\text{max}} \approx 4.21 \times 10^{21}$ photons/m$^2$/s/$\mu$m. 

\begin{figure}[ht!]
\includegraphics[width=13.8cm]{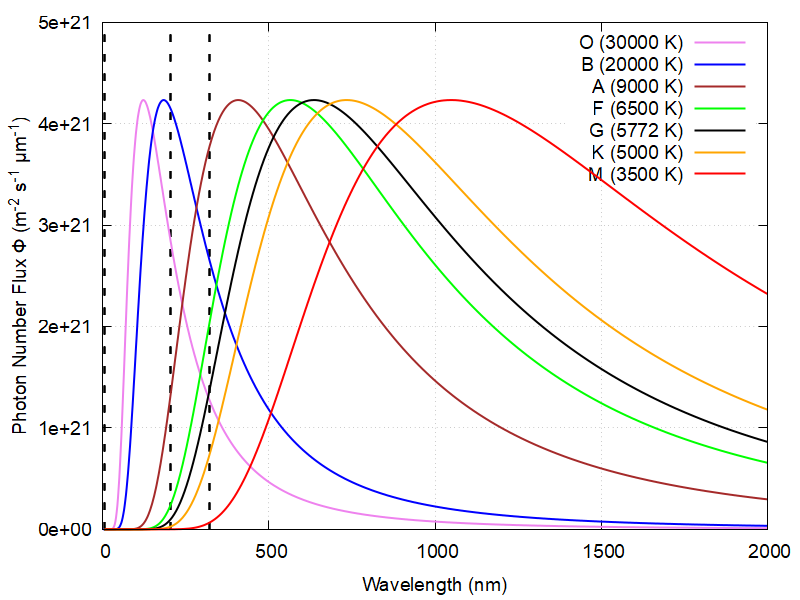}
\caption{The photon fluxes as a function of wavelength of different star types at the top of the atmosphere of a planet like Earth at the distance of planet from star that normalizes the wavelength integrated spectral constant to that of the Sun (1366 W/m$^2$). The vertical dashed lines mark the wavelength limits of the hard (ionizing) UV-C radiation (10-205 nm) and soft (dissipative structuring) radiation (205-320 nm).}
\label{fig:Spectral_Photon_Flux}
\end{figure}

In figure \ref{fig:Spectral_Photon_Flux} the spectral photon flux $\Phi(\lambda_{\text{nm}})$ (photons/m$^2$/s/$\mu$m) is plotted over the wavelength region 0--2000 nm for all star types. Identical photon flux at maximum across all star types is expected due to the normalization of integrated energy flux to 1366 W/m². The scaling of $(R/d)^2 \propto 1/T^4$ and the photon flux's $\lambda^{-4}$ dependence cancel out the temperature effects at the peak.

\begin{figure}[ht!]
\includegraphics[width=13.8cm]{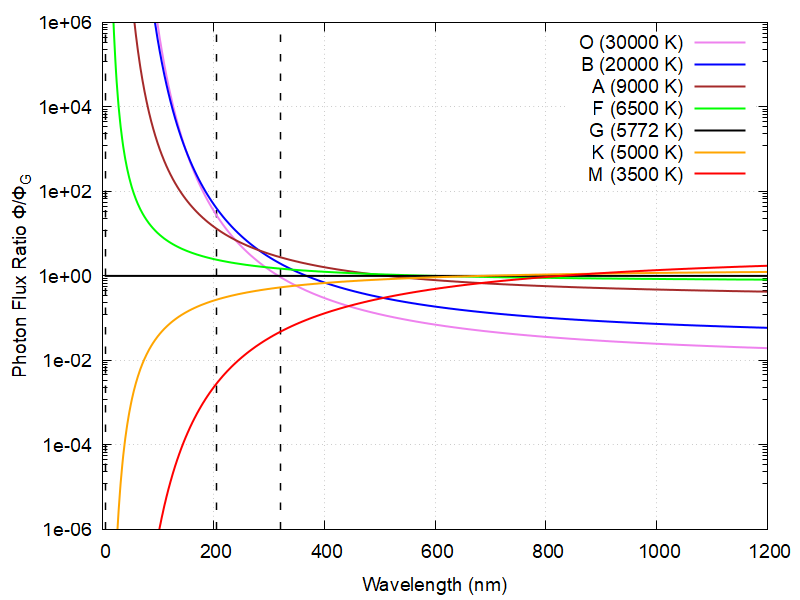}
\caption{Ratio of photon number fluxes as a function of wavelength for all star types compared to a G type star, like our Sun, at the top of the atmosphere for a planet like Earth at the distance of planet from star that normalizes the wavelength integrated spectral constant to that of the Sun (1366 W/m$^2$). The vertical dashed lines mark the wavelength limits of the hard (ionizing) UV-C radiation (10-205 nm) and soft (dissipative structuring) radiation (205-320 nm).}
\label{fig:Photon_flux_ratio}
\end{figure}

Figure \ref{fig:Photon_flux_ratio} plots the ratio of the photon number fluxes as a function of wavelength for all star types compared to a G-type star, like our Sun, at the top of the atmosphere for a planet like Earth at the distance of planet from star that normalizes the wavelength integrated spectral  constant to that of the Sun (1366 W/m$^2$). 

\section{Photon fluxes in Dissipative Production ($P$) and Ionization/Degradation ($D$) Regions}

For determining the viability of photochemical dissipative structuring on a given star type, spectral photon fluxes, rather than energy fluxes, are most relevant. Table \ref{tab:Photon_fluxes_top_atm} lists the integrated continuous photon fluxes at the top of the planet's atmosphere with the planet--star distance normalized to the solar constant. The spectral regions considered are; ionization/degradation, $D$ (10--205 nm), dissipative production, $P$ (205--320 nm) and HCN precursor production, $H$ (120-146 nm). The shortest wavelength of the destructive ionization region is taken to be 10 nm since there are few photons of shorter wavelength than this for all star types. The upper limit for the dissipative structuring regime is taken to be 320 nm corresponding to 3.9 eV since longer wavelengths do not have sufficient energy for transforming single covalent bonds between organic atoms.

HCN, an important precursor for the dissipative structuring of the nucleobases \cite{Michaelian2021}, can be produced in the upper atmosphere from the photon dissociation (120--146 nm) of N$_2$ and a carbon source (e.g., CH$_4$). The HCN produced would be protected once in the lower atmosphere since the gas-phase dissociation energy of the C$\equiv$N triple bond of HCN is $\approx 744$ kJ/mol (7.71 eV or 161 nm).

\begin{table}[ht!]
\centering
\caption{Integrated energy flux (W/m$^2$) and photon number fluxes (ph/m$^2$/s) at the top of the atmosphere of planets orbiting main-sequence stars at distances normalized to the solar constant (1366 W/m$^2$). Wavelength ranges are $D$: 10--205 nm (ionization dissociation), $P$: 205--320 nm (dissipative production), $H$: 120--146 nm (HCN production), and the ratio $P/D$, based on integration of the photon fluxes under the curves of Figure~\ref{fig:Spectral_Photon_Flux}.}
\label{tab:Photon_fluxes_top_atm}
\small
\begin{tabular}{l c c c c c c}
\toprule
Star & $T$ (K) & Energy Flux (W/m$^2$) & $D$ (ph/m$^2$/s) & $P$ (ph/m$^2$/s) & $H$ (ph/m$^2$/s) & Ratio $P/D$ \\
\midrule
O7 V & 30000 & 1366 & $6.107\times 10^{20}$ & $2.677\times 10^{20}$ & $1.288\times 10^{20}$ & 0.438 \\
B2 V & 20000 & 1366 & $4.865\times 10^{20}$ & $4.714\times 10^{20}$ & $1.040\times 10^{20}$ & 0.969 \\
A2 V & 9000 & 1366 & $5.321\times 10^{19}$ & $3.660\times 10^{20}$ & $3.558\times 10^{18}$ & 6.880 \\
F5 V & 6500 & 1366 & $6.499\times 10^{18}$ & $1.390\times 10^{20}$ & $1.427\times 10^{17}$ & 21.39 \\
G2 V & 5772 & 1366 & $2.319\times 10^{18}$ & $8.165\times 10^{19}$ & $2.998\times 10^{16}$ & 35.20 \\
K2 V & 5000 & 1366 & $5.309\times 10^{17}$ & $3.688\times 10^{19}$ & $3.268\times 10^{15}$ & 69.47 \\
M2 V & 3500 & 1366 & $3.550\times 10^{15}$ & $2.142\times 10^{18}$ & $1.945\times 10^{12}$ & 603.5 \\
\bottomrule
\end{tabular}
\end{table}

Earth's early Archean atmosphere was described in Section \ref{sec:EarlyEarth}. Below 145 nm, N$_2$ absorbs strongly (peak at 98 nm), with CO$_2$ absorbing strongly between 120--170 nm (peak at 126 nm), and H$_2$S absorbing between 190--200 nm (peak at 195 nm). Aldehydes, such as formaldehyde, which are common photochemical products on a 
CO$_2$ atmosphere absorb strongly between 285--310 nm \cite{Sagan1973}  , roughly corresponding to the UV-B region. 

\begin{figure}[ht!]
\includegraphics[width=13.8cm]{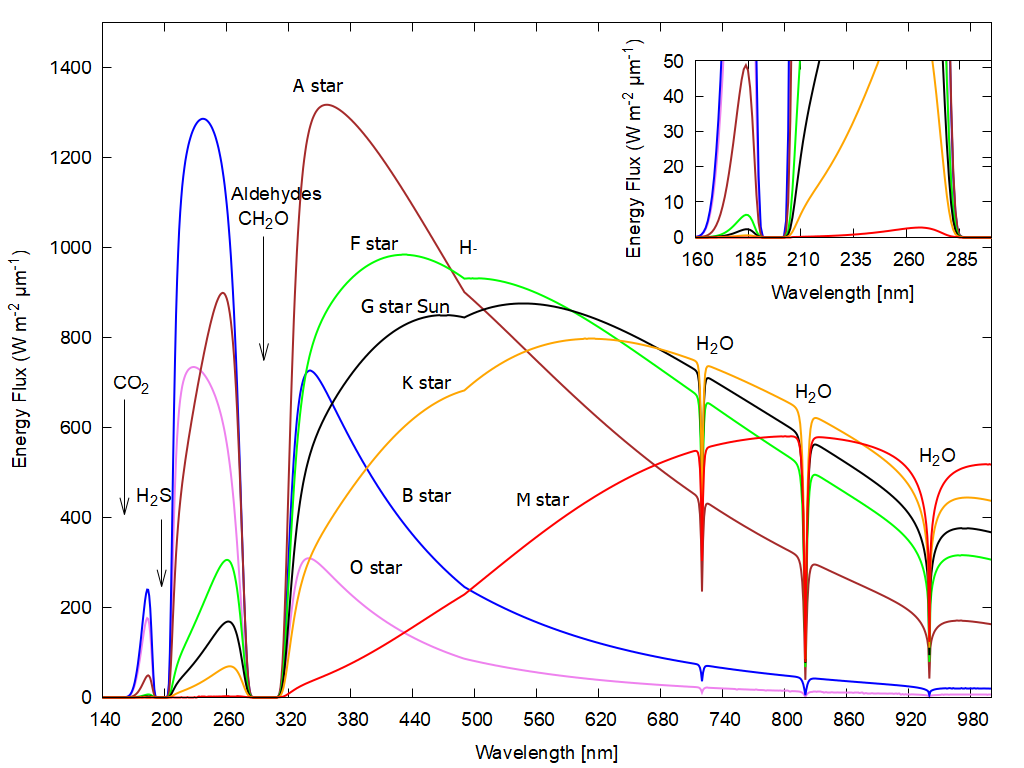}
\caption{The energy fluxes of different star types at the surface of a planet like early Earth at a distance from its star that normalizes the wavelength integrated spectral constant to that of the Sun (1366 W/m$^2$) as a function of wavelength. The inset is an amplification of the UV-C region.}
\label{fig:SpectraAtSurface}
\end{figure}

Modeling the absorption of these Archean atmospheric gases as Gaussian distributions centered on their peak absorption wavelength and with an extinction coefficient derived from their absorption cross section at maximum, the surface spectrum shown in Figure \ref{fig:SpectraAtSurface} is obtained. Integration over the region 240-290 nm for our Sun at zenith gives a surface energy flux of $\sim5$ Wm$^{-2}$, which is consistent with estimates by Sagan \cite{Sagan1973}.

Table \ref{tab:Photon_fluxes_Earth_surf} lists the values of the total integrated energy flux at the planet surface as well as the number of photons in the degradation wavelength region $D$ (10--205 nm), dissipative production region $P$ (205--320 nm), and HCN production region $H$ (120-146 nm), and the ratio $P/D$. For the larger stars, having a greater proportion of energy emitted at short wavelength, most of the incident energy is absorbed in the atmosphere of the planet and does not reach the surface.

\begin{table}[ht!]
\centering
\caption{Integrated energy flux (W/m$^2$) and photon number fluxes (ph/m$^2$/s) at the surface of an Earth-like planet orbiting main-sequence stars at distances normalized to a top-of-atmosphere solar constant of 1366 W/m$^2$. Wavelength ranges are $D$: 10--205 nm (ionization destruction), $P$: 205--320 nm (dissipative production), $H$: 120--146 nm (HCN production), and the ratio $P/D$, based on integration under the curves of Figure~\ref{fig:SpectraAtSurface}.}
\label{tab:Photon_fluxes_Earth_surf}
\small
\begin{tabular}{l c c c c c c}
\toprule
Star & $T$ (K) & Energy Flux (W/m$^2$) & $D$ (ph/m$^2$/s) & $P$ (ph/m$^2$/s) & $H$ (ph/m$^2$/s) & Ratio $P/D$ \\
\midrule
O7 V & 30000 & 91.57 & $1.930\times 10^{18}$ & $5.099\times 10^{19}$ & $9.945\times 10^{11}$ & 26.42 \\
B2 V & 20000 & 208.1 & $2.596\times 10^{18}$ & $9.251\times 10^{19}$ & $6.875\times 10^{11}$ & 35.64 \\
A2 V & 9000 & 522.9 & $4.999\times 10^{17}$ & $6.040\times 10^{19}$ & $1.134\times 10^{10}$ & 120.8 \\
F5 V & 6500 & 602.8 & $6.301\times 10^{16}$ & $1.885\times 10^{19}$ & $2.534\times 10^{8}$ & 299.1 \\
G2 V & 5772 & 618.9 & $2.199\times 10^{16}$ & $1.004\times 10^{19}$ & $4.020\times 10^{7}$ & 456.7 \\
K2 V & 5000 & 629.7 & $4.771\times 10^{15}$ & $3.954\times 10^{18}$ & $2.931\times 10^{6}$ & 828.8 \\
M2 V & 3500 & 631.0 & $2.436\times 10^{13}$ & $1.504\times 10^{17}$ & $4.417\times 10^{2}$ & 6174 \\
\bottomrule
\end{tabular}
\end{table}

\section{Net Molecular Production Rates}
\label{sec:MolecularProduction}
\subsection{Dynamical Equations}

Our analysis considers a photochemical dissipative structuring linear sequence requiring 6 independent photon absorption steps to produce the final molecule M (loosely based on the photochemical dissipative synthesis of adenine \cite{Michaelian2021}) starting with an initial constant concentration [M$_0$] of the precursor (e.g., HCN for adenine). For each step, M$_{i-1}$ + photon (flux $P$) $\to$ M$_i$ with rate $\alpha P$ [M$_{i-1}$] (assuming the same photochemical cross section for production, $\alpha$, for all 6 steps). For each intermediate molecule M$_i$ (i=1 to 5) and final M (i=6), the rate of photochemical destruction by ionizing radiation of flux $D$ is $\beta D$[M$_i$]. 

Photochemically dissipatively structured molecules not only have to survive dissociation or degradation by ionizing radiation, but also natural chemical destructive processes occurring at finite temperature in water solvent (e.g., hydrolysis, deamination or radical oxidation) with rate constant $k$, also assumed to be the same for all intermediates. Such destructive processes acting against molecular build-up are often neglected in photochemical feedstock scenarios but will be shown here to be critically important to discriminating between star types for abiogenesis since destructive chemical processes can overwhelm photochemical production on low mass stars.

The differential equations describing the concentration dynamics are thus,

\begin{eqnarray}
\frac{d[\text{M}_1]}{dt} &= \alpha P [\text{M}_0] - (\alpha P + \beta D + k ) [\text{M}_1], \nonumber \\
\frac{d[\text{M}_2]}{dt} &= \alpha P [\text{M}_1] - (\alpha P + \beta D+ k ) [\text{M}_2], \nonumber \\
&\vdots \nonumber \\
\frac{d[\text{M}_5]}{dt} &= \alpha P [\text{M}_4] - (\alpha P + \beta D+ k ) [\text{M}_5], \nonumber \\
\frac{d[\text{M}]}{dt} &= \alpha P [\text{M}_5] - (\beta D + k) [\text{M}] \label{eq:ProdRate}.
\end{eqnarray}

The initial concentration for the precursor is taken to be [M$_0$](0) = $6\times 10^{-5}$ M (e.g., HCN, see reference \cite{Michaelian2021}), and the intermediates [M$_i$](0) = 0 for i=1 to 5. 

Most hydrolysis rate constants at 85 $^\circ$C for small molecules are between $1\times 10^{-7}$ s$^{-1}$ and $1\times 10^{-6}$ s$^{-1}$. For example, for HCN, formamide, and AICN (molecules on route to adenine) they are $5.13\times 10^{-7}$, $6.69\times 10^{-7}$ and $2.94\times 10^{-7}$ respectively (see reference \cite{Michaelian2021} and references therein). A study on adenosine hydrolysis (glycosidic bond cleavage between adenine and ribose) at pH 7 \cite{BorquezEtAl2005}, extrapolated to 85$^\circ$C (early Archean surface temperature \cite{KastingAckerman1986, Kasting1993, KnauthLowe2003}) via an Arrhenius plot from measurements at 110--190$^\circ$C gives, $k = 1.038 \pm 0.360 \times 10^{-8} \, \text{s}^{-1}$. This corresponds to a half-life of approximately 2.12 years (activation parameters $\Delta H^\ddagger = 28.0$ kcal/mol and $\Delta G^\ddagger = 32.9 \pm 0.3$ kcal/mol). Any other chemical reactions (e.g., deamination, oxidation) will only increase these rates of degradation. For example, for adenine residues in DNA, deamination to hypoxanthine occurs at pH 7.6, 110 $^\circ$C with a rate constant of $k = 4 \times 10^{-8} \, \text{s}^{-1}$ \cite{KarranLindahl1980}. For ribose at 85 °C the rate constant for enolization/isomerization to other pentoses (e.g., arabinose, xylose), followed by further dehydration, is large, $k\sim 3.7\times 10^{-5}$ s$^{-1}$ range (half-life of hours) \cite{LarraldeEtAl1995}. For more complex molecules, the rate constants can be similarly larger. For example, for RNA phosphodiester backbone cleavage rates in these warm conditions are typically $k \approx 10^{-5}  $ to $  10^{-6}  $ s$  ^{-1} $ for small oligomers (half-lives of hours to days) \cite{LiBreaker1999}. Furthermore, hard UV-C ($< 205$ nm) radiation reaching the surface could cause greater production of free radicals in water, leading to greater molecular degradation through oxidation.

The nominal value of the chemical degradation rate constant $k$ is therefore assigned to $k = 1\times 10^{-6} \, \text{s}^{-1}$ ($t_{1/2} \approx 8$ days) for each photochemical step, but results are also obtained for $k$ both an order of magnitude greater and lesser.

The cross sections for photochemical production $\alpha$ and photochemical destruction $\beta$ are given by,
$$ \alpha = q_P \cdot \sigma_P \text{ and } \beta = q_D \cdot \sigma_D$$
where $q_P$ and $q_D$ are the quantum efficiencies for the photochemical reaction, and  $\sigma_P$ and $\sigma_D$ are the molecular photon absorption cross sections which are related to the extinction coefficient \(\epsilon\) (M$^{-1}$ cm$^{-1}$) by,
\begin{equation}
\sigma = \frac{\epsilon \cdot \ln(10) \cdot 10^3}{N_A} \ \text{cm$^2$/molecule}
\label{eq:sigma}
\end{equation}
where Avogadro's number \(N_A = 6.022 \times 10^{23}\) molecules/mol, \(\ln(10) \approx 2.3026\), and the factor \(10^3\) converts L to cm\(^3\).

Values of the quantum yields $q$ in the case of the dissipative structuring of adenine, range from about $5\times10^{-3}$ to $5\times10^{-1}$ with the smaller values associated to molecular transformations through high energy barriers and the larger values associated to charge transfer reactions \cite{Michaelian2021}. The rate of the overall photochemical process will be mostly influenced by the smallest quantum yields (bottleneck in the sequential linear chain as defined by Equations (\ref{eq:ProdRate})). Therefore, being conservative, we assume the value of $q_P = 0.005$ for each of the 6 photochemical reactions. 

The extinction coefficients at maximum absorption $\epsilon$ for the intermediate molecules on route to adenine range from about $10^1$ to $10^4$ M$^{-1}$ cm$^{-1}$ (Tables 1 and 2 of reference  \cite{Michaelian2021}) and similarly for chlorophyll (Table 23 of reference \cite{MichaelianSimeonov2026}) . For our analysis based only loosely on the photochemical route to adenine, we remain conservative and choose a value of $\epsilon = 2,000$ M$^{-1}$ cm$^{-1}$ for all 6 photochemical reactions. Eq. (\ref{eq:sigma}) then gives $\sigma_P=1.274\times 10^{-18}$ cm$^2$/molecule at wavelength of maximum absorption. For comparison, the experimentally determined maximum absorption cross sections for two molecules involved in the dissipative structuring of adenine are; formamide $2.29 \times 10^{-19}\ \text{cm²/molecule}$ at 220 nm and  AICN $4.09 \times 10^{-17}\ \text{cm²/molecule}$ at 250 nm \cite{KochRodehorst1974}.

The calculation of the photochemical reaction cross section $\alpha$ requires a convolution over the $P$ wavelength region,
\begin{equation}
\alpha P = \int_{205}^{320} q_P(\lambda) \cdot \sigma_P(\lambda) \cdot P(\lambda) d\lambda \approx \frac{1}{6} \cdot q_P^{max}\cdot \sigma_P^{max}\cdot P,\\
\label{eq:Approx}
\end{equation}
where $q_P(\lambda), \sigma_P(\lambda)$, and $P(\lambda)$ are the quantum yields, absorption cross sections, and photon intensities, respectively, as functions of wavelength. However, the first two of these have not yet been determined for any of the photochemical reactions leading to the fundamental molecules. In reference \cite{Michaelian2021} for the dissipative structuring of adenine, we assumed that the cross sections and quantum yields were at maximum value in a window around their peak of $\pm 10$ nm and zero elsewhere. This gives the factor of 20 nm/(320-205 nm) $\approx 1/6$ in the right hand side of Equation (\ref{eq:Approx}). Therefore,

\begin{equation}
\alpha \approx \frac{1}{6} \cdot \frac{q_P^{max} \cdot \sigma_P^{max}}{10^4} = 6.372 \times 10^{-25}\ \text{m²/molecule},\\
\end{equation}
where the factor of $10^4$ converts cm$^2$ to m$^2$ for use with the photon fluxes per m$^2$ ($P$ and $D$ of Table \ref{tab:Photon_fluxes_Earth_surf}). 
 
$\beta$, the cross section for photochemical degradation in the $D$ region, can then be  determined as the value which gives the correct concentration of adenine at 30 Archean days, $[\text{M}]$(30 days) = $5\times 10^{-8}$ M, obtained in a detailed calculation including all relevant photochemical and chemical reactions (Figures 10 and 11 of reference \cite{Michaelian2021}). The cross sections for photochemical degradation obtained in this manner, using the $P$ and $D$ values for the Sun, is found, using Equations (\ref{eq:ProdRate}), to be, 
\begin{equation}
\beta=4.843\times 10^{-22}\ \text{m²/molecule}.
\end{equation}

\subsection{Results}
\label{sec:Results}

Equations (\ref{eq:ProdRate}) are solved numerically in time steps of one second using the values of photon fluxes $P$ and $D$ given in Table \ref{tab:Photon_fluxes_Earth_surf} for each star type during the 3.5 hour daylight period (assuming the star always at the zenith), and $P=0$ and $D=0$ during the 10.5 hour night period of a 14 hour Archean day. 

\begin{figure}[ht!]
\includegraphics[width=13.8cm]{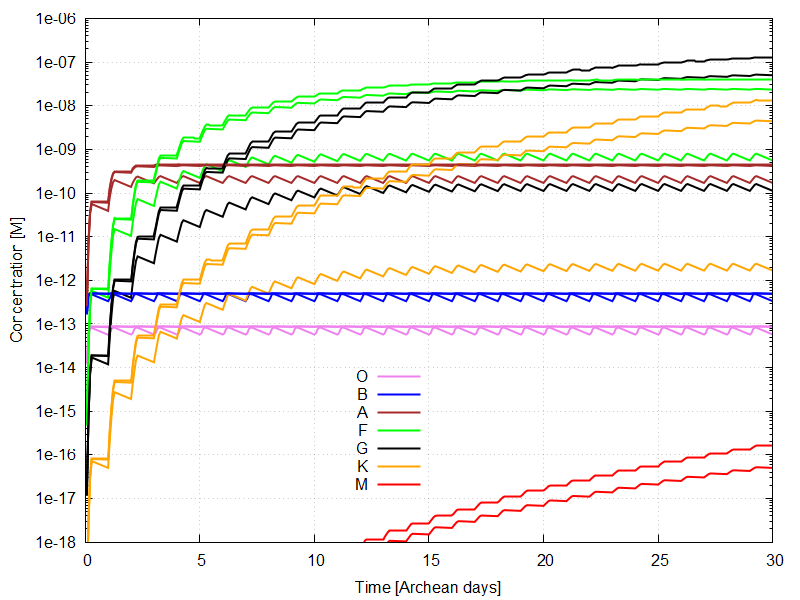}
\caption{The concentration of a fundamental molecule requiring 6 independent photochemical reactions (e.g., photochemical production of adenine \cite{Michaelian2021}) as a function of time in Archean days (3.5 hours of solar zenith light and 10.5 hours of darkness) on planets of different star types (different colors) at distances from their star normalized to the solar constant. The plots assume photochemical production cross sections (dissipative structuring) $\alpha=6.372\times 10^{-25}$ m$^2$/photon and photodegradation cross sections $\beta = 4.843\times 10^{-22}$ m$^2$/photon. Results are plotted for three values of the chemical degradation rate constant $k$, the nominal value of $10^{-6}$ s$^{-1}$ (middle curve), $10^{-7}$ s$^{-1}$ (upper curve) and $10^{-5}$ s$^{-1}$ (lower curve).}
\label{fig:Molecule_Concentration_all_k}
\end{figure}

\begin{figure}[ht!]
\includegraphics[width=13.8cm]{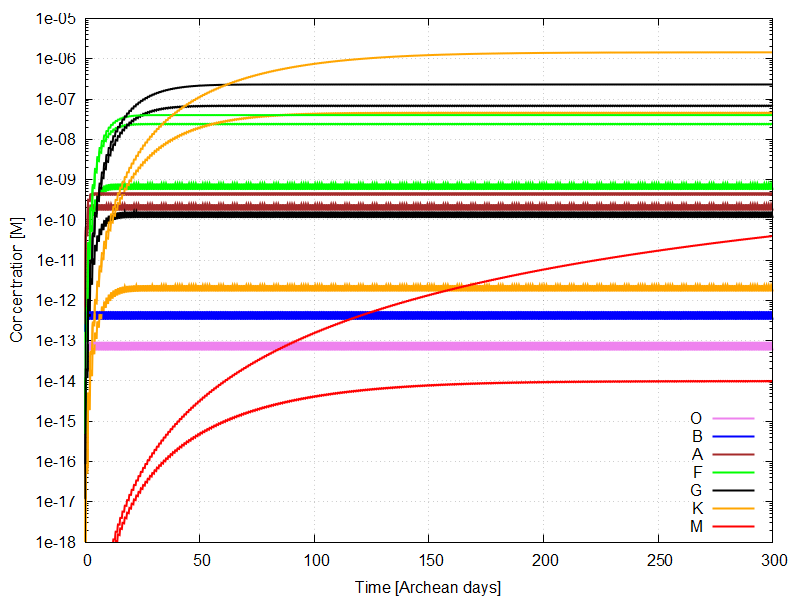}
\caption{The same as for Figure \ref{fig:Molecule_Concentration_all_k} but with the calculations extended out to 300 Archean days.}
\label{fig:Molecule_Concentration_all_k300d}
\end{figure}

Figure \ref{fig:Molecule_Concentration_all_k} shows the time dependence of the molecular concentrations over 30 Archean days for the three different  values of the chemical degradation rate constant $k$ for each star type. The graph shows a rapid rise to low concentrations, on the order of $10^{-13}$, $10^{-12}$ and $10^{-10}$ M for O, B and A-type stars respectively, and a slower rise to very low concentrations for M-type stars ($\sim4\times 10^{-17}$ M). Much higher concentrations ($\sim 10^{-7}$ M) are found for the G and F-type stars. The plots for the two smallest values of $k$ almost overlap, indicating that at these $k$ values chemical degradation has little effect on the concentrations for all stars types except K and M. Figure \ref{fig:Molecule_Concentration_all_k300d} plots the same concentrations out to 300 days. 

In the stationary (steady) state ($t \to \infty$), the concentrations are approximately,

\begin{equation}
[\text{M}]_{\infty} \approx [\text{M}_0]\left(\frac{0.25\alpha P}{0.25\alpha P + 0.25\beta D+ k}\right)^5 \frac{0.25\alpha P}{(0.25\beta D+k)},
\label{eq:Minfin}
\end{equation}
where the factor of 0.25 = 3.5/14 accounts the 3.5 hours of direct overhead sunlight in a 14 hour Archean day.

The time to reach 99\% of $[\text{M}]_\infty$ is approximated during the light period, where the system approaches $[\text{M}]_\infty$. The solution for $[\text{M}]$ is complicated due to the sequential reactions, but the rate-limiting step is governed by $\beta D + k$. Thus:
\begin{equation}
[\text{M}](t) \approx [\text{M}]_\infty \left(1 - e^{-(\beta D + k) t}\right),
\end{equation}
and setting $[\text{M}](t) = 0.99 [\text{M}]_\infty$ gives,
\begin{equation}
1 - e^{-(\beta D + k) t} \approx 0.99,
\end{equation}
or,
\begin{equation}
t \approx \frac{\ln(100)}{\beta D + k} \approx \frac{4.60517}{\beta D + k}.
\end{equation}
Since the light period is 3.5/14 of the day (50400 s), the time in Archean days ($t_{d}$) to reach 99\% of $[\text{M}]_\infty$ is,
\begin{equation}
 t_{d} = \frac{14}{3.5\cdot 50400}\cdot t \approx \frac{4.60517}{(\beta D + k) \cdot 12600}.
\end{equation}

Table \ref{tab:ConcentrationTimes} lists the stationary state concentrations and the time to reach 99\% of this value for each star type given the production $P$ and degradation $D$ photon fluxes and the three chemical degradation $k$ values.

\begin{table}[ht!]
\centering
\caption{Stationary concentrations $[M]_\infty$ (mol/L) and time to 99\% $[M]_\infty$ (Archean days) for planets on different star types, with $[M_0] = 6.0 \times 10^{-5}$ mol/L, $\alpha=6.372\times 10^{-25}$ m$^2$/photon, $\beta = 4.843\times 10^{-22}$ m$^2$/photon, 14-hour Archean day (3.5 hours light, 10.5 hours dark), and for the three $k$ (s$^{-1}$) values. See Section \ref{sec:M-Flare} for the determination of values for the flaring M star.}
\small
\setlength{\tabcolsep}{3.5pt}
\begin{tabular}{l c c c c c c c c}
\toprule
& & & \multicolumn{2}{c}{$k=10^{-7}$} & \multicolumn{2}{c}{$k=10^{-6}$} & \multicolumn{2}{c}{$k=10^{-5}$} \\
\cmidrule(lr){4-5} \cmidrule(lr){6-7} \cmidrule(lr){8-9}
Star & $P$ & $D$ & {$[M]_{\infty}$} & {Time} & {$[M]_{\infty}$} & {Time} & {$[M]_{\infty}$} & {Time} \\
& ph/m$^{2}$/s & ph/m$^{2}$/s & (M) & (days) & (M) & (days) & (M) & (days) \\
\midrule
O7 V & $5.099\times 10^{19}$ & $1.930\times 10^{18}$ & $8.903\times 10^{-14}$ & 0.390 & $8.706\times 10^{-14}$ & 0.390 & $6.989\times 10^{-14}$ & 0.390 \\
B2 V & $9.251\times 10^{19}$ & $2.596\times 10^{18}$ & $5.061\times 10^{-13}$ & 0.290 & $4.979\times 10^{-13}$ & 0.290 & $4.231\times 10^{-13}$ & 0.290 \\
A2 V & $6.040\times 10^{19}$ & $4.999\times 10^{17}$ & $4.592\times 10^{-10}$ & 1.51 & $4.246\times 10^{-10}$ & 1.50 & $2.042\times 10^{-10}$ & 1.45 \\
F5 V & $1.885\times 10^{19}$ & $6.301\times 10^{16}$ & $3.999\times 10^{-8}$ & 11.9 & $2.395\times 10^{-8}$ & 11.6 & $6.677\times 10^{-10}$ & 9.02 \\
G2 V & $1.004\times 10^{19}$ & $2.199\times 10^{16}$ & $2.303\times 10^{-7}$ & 34.0 & $6.800\times 10^{-8}$ & 31.4 & $1.345\times 10^{-10}$ & 17.7 \\
K2 V & $3.954\times 10^{18}$ & $4.771\times 10^{15}$ & $1.447\times 10^{-6}$ & 152 & $4.532\times 10^{-8}$ & 110 & $2.004\times 10^{-12}$ & 29.7 \\
M2 V & $1.504\times 10^{17}$ & $2.436\times 10^{13}$ & $3.350\times 10^{-9}$ & 3270 & $9.913\times 10^{-15}$ & 361 & $1.120\times 10^{-20}$ & 36.5 \\
M-flare & $2.256\times 10^{17}$ & $3.842\times 10^{13}$ & $2.786\times 10^{-8}$ & 3650 & $1.084\times 10^{-13}$ & 365 & $1.270\times 10^{-19}$ & 36.6 \\
\bottomrule
\end{tabular}
\label{tab:ConcentrationTimes}
\end{table}

\begin{figure}[ht!]
\includegraphics[width=13.8cm]{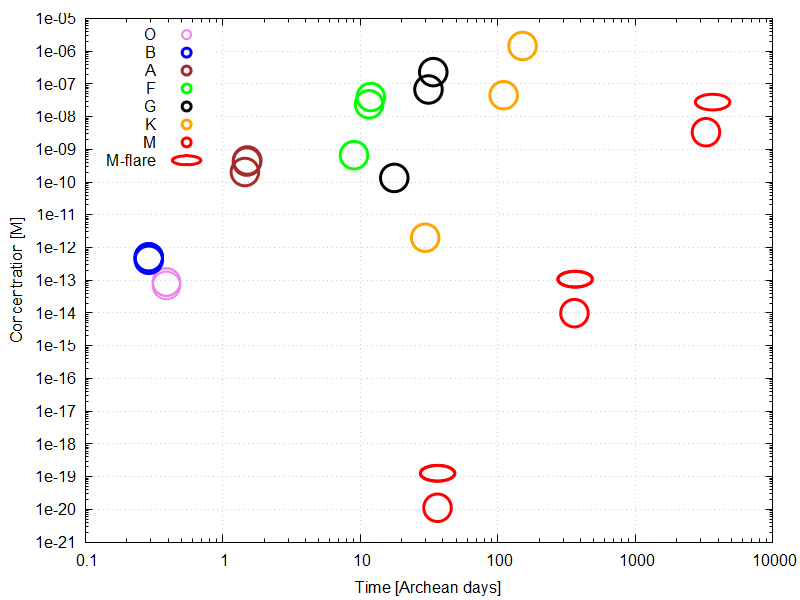}
\caption{Stationary state values of the concentrations of fundamental molecules versus the time in days to reach 99\% of this value for each star type and for the three chemical degradation rate constants $k$, of $10^{-7}$, $10^{-6}$ and $10^{-5}$ (giving highest to lowest stationary state concentrations, respectively). For O and B-type stars, all three $k$ values give similar results, indicating that chemical degradation has little effect on these stationary state molecular concentrations, while chemical degradation has a great effect on M stars. The results for highly active flaring M stars (see Section \ref{sec:M-Flare}) are given by the red ellipses. Both axises are logarithmic.}
\label{fig:Stat_con_time}
\end{figure}

Figure \ref{fig:Stat_con_time} plots the stationary state values of the concentrations versus the time in Archean days to reach 99\% of the stationary state value for all star types and for the three $k$ values (representing hydrolysis, deamination, oxidation, or other chemically mediated degradation). For the nominal chemical degradation rate constant of $k=10^{-6}$, G-type stars give the highest molecular concentrations ($6.80\times 10^{-8}$ M), with somewhat smaller values for F and K-type stars ($2.40\times 10^{-8}$ and $4.53\times 10^{-8}$ M, respectively). 

For the nominal value of $k$, the time to reach these stationary values is 11.6, 31.4 and 110 Archean days for F, G and K star types, respectively (see Table \ref{tab:ConcentrationTimes} and Figure \ref{fig:Stat_con_time}). O and B stars produce stationary concentrations about 6 orders of magnitude smaller than G-type stars, while M stars produce stationary concentrations 7 orders of magnitude smaller, and it takes much longer (361 Archean days) to reach this very low stationary concentration. Highly active flaring M stars (see Section \ref{sec:M-Flare}) give concentrations only about an order of magnitude larger than quiescent M stars, and take about the same time to reach those values.

The stationary-state concentrations of fundamental biomolecules are a key determinant of whether photon-driven dissipative structuring—leading to life as we know it—can occur on planets orbiting other stellar types. Extremely low concentrations would hinder polymerization into functional macromolecules such as RNA, DNA, and proteins. Equally critical is the timescale required to reach these stationary levels; if too long, periodic catastrophic events (e.g., large impactors or intense hard UV-C flares) could repeatedly reset molecular inventories to near-zero values. The influence of such disruptive events—and their frequency as a function of spectral type—is examined in the following subsection, along with other factors governing net molecular accumulation.

\subsection{Other Factors Relevant to Molecular Buildup}

\subsubsection{HCN Precursor Concentrations}

Stationary state molecular concentrations depend linearly on the initial precursor concentrations (Eq. \ref{eq:Minfin}). For adenine and the other nucleobases, as well as for important amino acids, it is generally assumed that HCN or its derivatives (e.g., cyanogen) were the precursors. HCN stationary state concentrations in the sea surface microlayer were assumed to be constant ($[\text{M}_0] = 6\times 10^{-5}$ M \cite{Michaelian2021}) and the same for all planets on the different star types. 

Most of the HCN production on early Earth was through photochemical reactions in Earth's upper atmosphere \cite{Zahnle1986, TianEtAl2011} with production scaling linearly with the CH$_4$ or H$_2$ concentration \cite{PearceEtAl2022}. The integrated photon flux in the HCN production wavelength region $H$ (120-146 nm) for O and B-type stars have intensities 4 orders of magnitude greater than G-type stars, while quiescent M-type stars have intensities 4 orders of magnitude lower (Table \ref{tab:Photon_fluxes_top_atm}), implying initial precursor concentrations could be significantly different for planets on different star types. 

The atmospheric production of HCN, however, may have a more complex dependence on star-type. For example, young, very active M stars could have time-averaged intensities in the HCN production region $H$ (120-146 nm) up to two orders of magnitude larger than quiescent G stars \cite{LoydEtAl2018} although the dissociation energy of HCN corresponds to a photon wavelength of approximately 93 nm, which will also be of higher intensity for these stars. This dissociation wavelength is also exactly where O-type stars peak in photon intensity.

Precursor feedstock concentrations will thus be a complicated function of star type, atmospheric chemistry, and planetary dynamism \cite{AirapetianEtAl2016,  AirapetianEtAl2020, KobayashiEtAl2023}. Empirical data, or detailed 3D climate-chemistry models \cite{ChenEtAl2021}, will be needed to estimate precursor concentrations for a particular star-planet combination. However, since stationary state concentrations depend linearly on the initial precursor concentration ($\text{M}_0$, Eq. (\ref{eq:Minfin})), results could be scaled accordingly.

\subsubsection{Clouds and Hazes}
\label{sec:Hazes}

On Earth, the soft UV-C surface flux was relatively stable over the late Hadean and throughout the Archean \cite{MeixnerovaEtAl2021}, providing a continuous long term impressed generalized chemical potential for the accumulation of complexity through photochemical dissipative structuring. 

CO$_2$ is the main absorbing atmospheric species below about 205 nm. Atmospheric Rayleigh scattering (with a $1/\lambda^4$ dependence) also begins to become important below 205 nm). However, significant volcanic activity could lead to significant amounts of  H$_2$S and SO$_2$ in the atmosphere which would narrow the surface window from $>205$ nm to about $>230$ nm depending on the amount of these trace gases \cite{Sagan1973}. Organic hazes linked to CH$_4$ photochemistry with the sulfur gases could produce aggregates of elemental sulfur S$_8$ (other sulfur containing molecules are easily photolyzed by radiation between 200 and 300 nm) that can become optically thick in the soft UV-C region through absorption or Mie scattering \cite{GaoEtAl2017}, thereby decoupling surface UV from the stellar type.

On Venus, the very high CO$_2$ partial pressure, and the thick, globally encircling sulfuric acid (H$_2$SO$_4$) clouds result in strong Mie scattering, preventing both the hard and soft UV-C from reaching the surface. On Earth, water clouds reduce the transmission of both the hard and soft UV-C regions by similar amounts, typically 20-60\%, depending on cloud thickness, again mainly through Mie scattering \cite{KyllingEtAl1997}. 

If clouds or haze close the soft UV-C window, it effectively renders that planet surface spectrum adverse to TDTOL UV-C-driven abiogenesis, regardless of the host star type. However, the first steps of abiogenesis, the dissipative structuring of strongly absorbing UV-C organic pigments \cite{Michaelian2026b}, may still occur in the atmosphere (e.g., the still uncharacterized UV absorbers in the upper cloud deck of Venus \cite{MolaverdikhaniEtAl2012} and benzene and other organics in the higher atmosphere of Titan \cite{CapalboEtAl2016, TribbettEtAl2021}).

\subsubsection{Flaring Stars}
\label{sec:M-Flare}

Studies have concluded that flaring M stars may be potentially viable for a UV photochemical feedstock scenarios \cite{RanjanSasselov2016, RimmerEtAl2021, PaudelEtAl2024}. From our non-equilibrium thermodynamic perspective, a more relevant question may be whether flaring M stars could sustain UV-C dissipative structuring (molecular concentration build-up, complexation through association, replication, etc.) while under increased photochemical degradation.

On average, M0-M3 dwarfs flare 2.1\%, while M4-M6 5\%, of the time \cite{PaudelEtAl2024}. The enhancements over quiescent levels during a flare in the NUV region (170–320 nm) are mostly less than a factor of 10 (Figure 12 of reference \cite{PaudelEtAl2024}). During M star flares, there is a somewhat larger increase in the hard UV-C ($<205$ nm - $D$ value, Table \ref{tab:Photon_fluxes_Earth_surf}) region compared to the soft UV-C + UV-B (205–320 nm - $P$ value), giving a much smaller $P/D$ ratio (similar to an A star with a black-body temperature of around 9,000-10,000 K \cite{PaudelEtAl2024}). 

Assuming a very active flaring scenario of a $5\times 10 =50$\% increase in the time averaged value of the soft UV-C flux $P$ for a very active M star gives a time averaged $(P/D)_{avg}$ ratio of,
\begin{equation}
(P/D)_{avg} = 0.05(P/D)_{A} + 0.95(P/D)_{M} = 5871,
\end{equation}
where $(P/D)_{A} = 120.8$ for an A type star and $(P/D)_{M} = 6174$ for a M type star (Table \ref{tab:Photon_fluxes_Earth_surf}). A 50\% larger time averaged value of $P$ for the flaring star will be $P = 1.5\times 1.504\times 10^{17}=2.256\times 10^{17}$ ph/m$^2$/s, giving the time averaged value of $D=P/5871 = 3.842\times 10^{13}$ ph/m$^2$/s, about 58\% larger than nominal value for a quiescent M star (Table \ref{tab:Photon_fluxes_Earth_surf}).

Using these values in equation (\ref{eq:Minfin}) gives fundamental molecule stationary state concentration values of $2.786\times 10^{-8}$, $1.084\times 10^{-13}$, $1.270\times 10^{-19}$ M for the three different chemical degradation rate constants ($k=10^{-7}, 10^{-6}, 10^{-5}$ s$^{-1}$), each about an order of magnitude larger than for a quiescent M star (Table \ref{tab:ConcentrationTimes}). These values are plotted against the time required to reach the stationary state as the red ellipses in Figure \ref{fig:Stat_con_time}. For the nominal chemical degradation rate constant of $k=10^{-6}$ s$^{-1}$, the stationary state concentration of the dissipatively structured organic molecule for a very active flaring M star (flaring 5\% of the time with each and every flare providing 10 times the quiescent NUV flux) is still about 6 orders of magnitude smaller than that for a G-type star.

Although CO$_2$ or H$_2$ rich secondary atmospheres are not predicted for planets with surface temperatures $\le 500$ K \cite{HerbortEtAl2020}, assuming a much denser CO$_2$ or H$_2$ planetary atmosphere compared to Earth's early Archean atmosphere, shielding the surface completely from the hard UV-C (setting $D=0$ in Eq. (\ref{eq:Minfin})), does not change the concentrations notably for flaring low mass stars. This is because chemical degradation is much more important than photochemical degradation for low mass stars, simply because the time to reach the stationary state (for flaring and non-flaring stars) is very long (Figure \ref{fig:Stat_con_time}). This is not the case for high mass O to G stars for which photochemical degradation is more important than chemical degradation.

The transient and time random nature of occurrence of flares on low mass stars present adverse conditions for TDTOL mechanisms driving abiogenesis and evolution. Dissipative structures in general (e.g., molecular concentration profiles) are more stable under a continuous (relative to natural degradation times) impressed potential (i.e., constant UV-C light flux). For example, as mentioned previously, obtaining and maintaining high molar concentrations of the fundamental molecules with high Gibb's free energy in the stationary state, requires successive molecular transformations to occur, but predominantly in the forward direction, i.e., towards product concentration profiles which are increasingly dissipative after each step (Figure 3 of reference \cite{Michaelian2021}). This could not occur if the surface spectrum was rapidly varying in time. As another example, randomly timed, transient, intense soft UV-C fluxes are not conducive to procuring homochirality through the diurnal mechanism of UV-C and temperature assisted replication (UVTAR) of DNA or RNA \cite{Michaelian2018}.

Finally, active young M, and low mass K, dwarfs could also produce high O$_2$ environments on planets by dissociating atmospheric H$_2$O, CO or CO$_2$ \cite{LugerBarnes2015}, thereby leading to an abiotic ozone layer preventing the soft UV-C light from reaching the surface.

In conclusion, flaring M stars are most probably less likely to host dissipative structuring leading to abiogenesis and evolution on their planets than quiescent M stars.

\subsubsection{Transient X-ray and Far-Ultraviolet Outbursts of Main-Sequence Stars}

The average frequency of large ($>10^{32}$ erg) and small ($10^{30} - 10^{31}$ erg) X-ray (0.12–50 keV) or far-ultraviolet (FUV, 91–200 nm) outbursts is listed in Table \ref{tab:Outbursts} for main-sequence stars G2 V, K2 V, and M2 V, representing field stars ($\sim1-5$ Gyr). Large flares increase the flux by factors of $\sim10-100$, while small flares by factors of $\sim2-10$, lasting minutes to hours.

\begin{table}[h]
\centering
\caption{Frequency of large ($>10^{32}$ erg) and small ($10^{30}$--$10^{31}$ erg) X-ray (0.12--50 keV) and far-ultraviolet (FUV, 91--200 nm) outbursts for main-sequence G2 V, K2 V, and M2 V stars. Frequencies are the mean time in days between flares, derived from ROSAT, XMM-Newton, and GALEX data, with ranges reflecting variability in rotation and magnetic activity. Frequencies are higher for younger stars ($\sim$10--625 Myr) and lower for older stars ($>$5 Gyr). Complied with data from \cite{Gudel2004, Gudel2007, NamekataEtAl2022, NamekataEtAl2024a, NamekataEtAl2024b}.}
\label{tab:Outbursts}
\begin{tabular}{l c c}
\toprule
Star Type & Frequency of Large Outbursts (days) & Frequency of Small Outbursts (days) \\
\midrule
G2 V & 365--3650 & 0.2--2 \\
K2 V & 7--30     & 0.2--1 \\
M2 V & 3--30     & 0.05--0.2 \\
\bottomrule
\end{tabular}
\end{table}

As considered in the previous subsections, such X-ray and hard UV-C flares, under nominal early Archean atmospheric conditions, could negatively affect the concentration build-up of the precursors and intermediate molecules.  Extra atmospheric hard UV-C shielding will help only minimally in the dissipative structuring of the fundamental molecules since molecular degradation is mostly chemical, not photochemical, for low mass K and M stars.

Transient outbursts on higher-mass stars (O through G) are both far less frequent and substantially weaker in intensity because their convective zones and the dynamo processes that generate magnetic activity are buried much deeper beneath the photosphere.

\subsubsection{Coronal Mass Ejections and Stellar Energetic Particle Events}

Low-mass stars (K and M dwarfs) have convective envelopes that extend close to, or all the way to, the photosphere. Besides driving frequent and intense X-ray and FUV flares, this also drives associated coronal mass ejections and energetic ionized particle events \cite{AirapetianEtAl2020, NamekataEtAl2026} capable of eroding planetary atmospheres \cite{DongEtAl2018, Tabataba-VakiliEtAl2016, TilleyEtAl2019} thereby reducing  atmospheric pressures required to support standing bodies of liquid water and depleting atmospheric N$_2$ and CO$_2$ \cite{AirapetianEtAl2017, GarciaSageEtAl2017}, thus potentially exposing the surface to lethal flares of hard UV-C radiation. These events, particularly prominent for young low-mass stars, pose a severe obstacle for the dissipative structuring and persistence of an Earth-like early biosphere. 

\subsubsection{Stellar Lifetimes}

The more massive a star, the higher its internal temperature and therefore the greater its rate of nuclear burning, leading to shorter lifetimes (Table \ref{tab:StellarLifetimes}).

\begin{table}[h]
\centering
\caption{Average lifetimes in Gyr ($10^9$ years) for representative spectral types of main-sequence stars. Lifetimes are derived from stellar evolution models and scale approximately as $\tau \propto M^{-2.5}$, where $M$ is the stellar mass in solar masses. Ranges account for variations in mass and metallicity \cite{CoxPilachowski2000, LaughlinEtAl1997, Allen2000, SchallerEtAl1992}.}
\label{tab:StellarLifetimes}
\begin{tabular}{l c}
\toprule
Spectral Type & Lifetime (Gyr) \\
\midrule
O5 V  & 0.001--0.003 \\
O9 V  & 0.003--0.005 \\
B0 V  & 0.01--0.02 \\
B5 V  & 0.05--0.1 \\
A0 V  & 0.4--0.6 \\
A5 V  & 0.8--1.0 \\
F0 V  & 1.5--2.0 \\
F5 V  & 2.5--3.5 \\
G0 V  & 8--10 \\
G2 V  & 10 \\
G5 V  & 10--12 \\
K0 V  & 15--20 \\
K2 V  & 20--30 \\
K5 V  & 30--50 \\
M0 V  & 50--70 \\
M2 V  & 70--100 \\
M5 V  & 100--200 \\
\bottomrule
\end{tabular}
\end{table}

The earliest widely accepted evidence for microbial life on Earth dates to approximately 3.45\,Ga \cite{AllwoodEtAl2006}, roughly 1\,Gyr after the Sun reached the zero-age main sequence. If this $\sim$1\,Gyr interval is representative of the time required for carbon-based life to evolve to detectable microbial ecosystems, then stellar lifetimes (Table~\ref{tab:StellarLifetimes}) impose a strong additional constraint. O-, B-, and most A-type stars simply do not remain on the main sequence long enough for such biological development. Combined with their very low predicted concentrations of fundamental biomolecules due to photochemical degradation (Section~\ref{sec:Results}), this makes the presence of complex biology on planets of these high-mass stars improbable.

\subsubsection{Tidal Locking}

The dissipation of a planet's rotational energy through gravitational interaction with its star (and moons), implies that planets in habitable zones located close to their stars may become tidally locked, leaving one side in eternal starlight and the other in eternal darkness. 

For the first step of abiogenesis, the dissipative structuring of chromophores like the fundamental molecules of life, this is not necessarily an impediment. However, the TDTOL also asserts that more complex biosynthetic structuring during the Archean was dependent on the diurnal cycle of sunlight \cite{Michaelian2011}. For example, the enzyme-less light-induced denaturing during daylight and extension over night of RNA and DNA, and, as a consequence, life's homochirality, are predicted by TDTOL to require diurnal cycling of the UV-C light and high ocean surface temperatures (Ultraviolet and Temperature Assisted Replication - UVTAR \cite{MichaelianSantillan2019, Michaelian2018}). 

Planets on M and lower mass K stars in their habitable zone may become tidally locked within a few hundred million years, arresting evolution beyond the UV-C pigment epoch of dissipatively structured carbon based life.

\section{Relevant Biosignatures Under TDTOL}

If TDTOL correctly describes the origin of carbon-based life on Earth, a natural and physically grounded biosignature emerges: {\em the global planetary photon-dissipation rate} (or, equivalently, the global entropy production due to irreversible absorption and re-emission of stellar photons). According to TDTOL, dissipation begins in the soft UV-C + UV-B (205--320\,nm) and progressively broadens toward longer wavelengths as new pigments and biosynthetic pathways emerge incorporating energy storing molecules like ATP. More mature biospheres, characterized by greater photon absorption and biomass and tighter thermodynamic coupling between organisms and abiotic processes (e.g., water cycle, oceanic and atmospheric dynamics), exhibit correspondingly higher global entropy production \cite{Michaelian2012a, MichaelianCano2022}.

Detecting this thermodynamic biosignature requires measuring both the incident stellar spectrum and the planet’s emitted spectrum from the UV-C to far-infrared wavelengths, and from this computing the net entropy production using the Planck formula (see \citealt{Michaelian2012a} and \citealt{MichaelianCano2022} for detailed methodology). A significant excess of entropy production, above that expected for a lifeless world of the same albedo and temperature, would constitute strong evidence for a TDTOL-driven biosphere.

Emitted spectra of exoplanets are presently difficult to measure due to the saturating light of the star, particularly for the high mass stars, and also for the intermediate mass stars for which TDTOL predicts the most probable occurrence of abiogenesis through molecular dissipative structuring. A less challenging proxy measurement, therefore, may be the soft UV-C planet albedo. According to TDTOL, this should be low after the formation of a primordial UV-C ``pigment world'' \cite{Michaelian2024}, and would remain low throughout biological evolution of the planet. 

\begin{figure}[ht!]
\includegraphics[width=13.8cm]{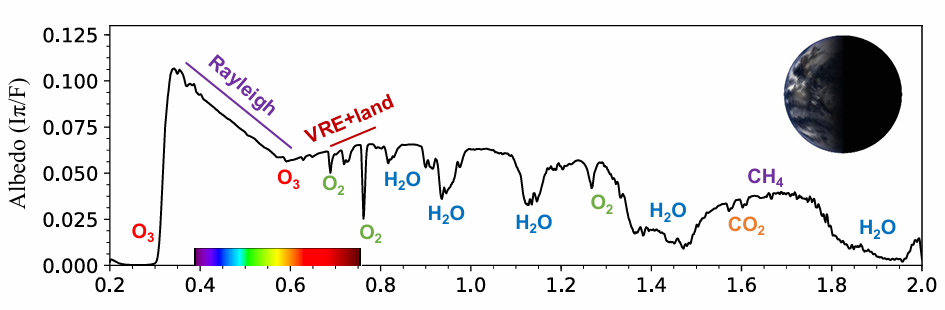}
\caption{The simulated wavelength dependent albedo of present day Earth \cite{SchwietermanEtAl2024} as a function of wavelength ($\mu$m). The soft UV-C albedo (205--285 nm) is very low, due principally to the absorption by ozone, which is considered under TDTOL to be a life synthesized UV-C pigment. It is argued that this region of the spectrum would show low planet albedo at the beginning of life, and would remain low throughout the entire evolution of life and, therefore, may be a more easily measured biosignature for dissipatively structured carbon based life on a planet at all stages of its evolution. Saturation effects due to Rayleigh scattering off atmospheric molecules with a $1/\lambda^4$ dependence is noted on the plot and would have to be accounted for in the soft UV-C region for planets with suspected early life which had not yet discovered oxygenic photosynthesis. The inset depicts the Earth at half illumination for which the simulation was performed. Reprinted from Schwieterman et al. \cite{SchwietermanEtAl2018, SchwietermanEtAl2024} under Creative Commons Attribution License CC-BY.}
\label{fig:Albedo}
\end{figure}

Figure~\ref{fig:Albedo} shows the wavelength-dependent albedo of present-day Earth simulated from measured atmospheric properties. Notably, the albedo in the soft UV-C range (205--285\,nm) is extremely low ($<0.05$). We have previously argued \cite{Michaelian2017} that this pronounced UV-C absorption, arising primarily from stratospheric ozone and molecular oxygen of biotic origin, can be regarded, within the TDTOL framework, as a continuation of biosphere dissipative structuring to the synthesis of the UV-C pigments oxygen and ozone. The possibility of considerable abiotic production of oxygen would have to be discarded.

During the Archean, in the absence of free oxygen and ozone, soft UV-C photons (205--285\,nm) would have reached Earth's surface essentially unattenuated (Figure~\ref{fig:Pigments}). There they would have been strongly absorbed by the fundamental biomolecules, acting as the primary UV-C pigments, yielding a correspondingly low planetary albedo in this wavelength region, similarly to what ozone does today (Figure~\ref{fig:Albedo}). Rayleigh scattering off atmospheric molecules with a $1/\lambda^4$ dependence introduces a modest wavelength-dependent offset, but this contribution can be accurately subtracted once the column densities of the dominant gases are constrained from other spectral regions. After such correction, a persistently low soft UV-C albedo may constitute a direct thermodynamic biosignature of an Archean-like biosphere commensurate with TDTOL.

\section{Discussion}

O-type stars produce high planetary surface photon fluxes in the ionization region ($< 205$ nm) for Archean Earth analogue atmospheres, leading to very low molecular steady state concentrations (on the order of $10^{-6}$ that of G-type stars for a nominal chemical degradation rate constant of $k=10^{-6}$ s$^{-1}$ - Table \ref{tab:ConcentrationTimes}). Furthermore, the dissociation energy of one of the most important precursors for Earth's fundamental molecules, HCN,  corresponds to a photon wavelength of approximately 93 nm, exactly where O-type stars peak in photon intensity.  These stars also have very short lifetimes, $< 0.005$ Ga, insufficient time for evolution to complex lifeforms such as bacteria, which took roughly 1 Gyr on Earth.

B-type stars suffer from the same afflictions as O-type stars, but to a somewhat lesser degree. For example, their dissipatively structured stationary state molecular concentrations for the nominal chemical degradation rate constant are more than five orders of magnitude smaller than that predicted for G-type stars (Table \ref{tab:ConcentrationTimes}). Their maximum lifetimes of $\sim 0.1$ Gyr again suggests low probability of finding bacteria on their planet surfaces.

A-type stars have stationary state molecular concentrations about two orders of magnitude lower than that of G-type stars and may live long enough $\sim1$ Gyr for the evolution to bacteria, but the greater than one order of magnitude higher hard UV-C ($< 205$ nm) radiation (Table \ref{tab:ConcentrationTimes}, column $D$) could cause other problems for incipient life, for example, the greater production of free radicals in water, requiring a different or thicker atmosphere than Earth's Archean to filter out the harmful hard UV-C light.

F and G-type stars appear to be the most adequate for hosting life like our own through photochemical dissipative structuring. They provide high stationary state concentrations of the fundamental molecules and arrive at these stationary state values within short time periods (a few weeks). They have low rates of X-ray, FUV and particle outbursts and they live long enough for the evolution of bacteria and even complex ecosystems. However, shorter stellar lifetimes of F type stars ($\sim 2$ Ga) imply that intelligent human-like civilizations probably only arise on G-type stars. F and G-type stars sum to only about 8-10\% of all stars in our galaxy \cite{CoxPilachowski2000}.

K-type stars suffer from a number of impediments to the dissipative structuring of life on their planets within habitable zones. First, although their the stationary state concentrations of the fundamental molecules are similar to that of G-type stars, they are more prone to natural periodic catastrophic events. Frequent outbursts of high intensity short wavelength X-rays and FUV, and high energy particles, with average periods of only days could degrade atmospheres and fundamental molecules formed in the soft UV-C. An order of magnitude of lower photon intensity in the HCN production $H$ region of the spectrum (Table \ref{tab:Photon_fluxes_top_atm}) would result in a further decrease in the stationary state concentrations. Tidal locking of planets on lower mass K-type stars would also mean no diurnal cycling necessary for UV-C induced RNA and DNA denaturing \cite{MichaelianSantillan2019} and homochirality \cite{Michaelian2018}, as well as night time extension (enzyme-less RNA/DNA replication) required for subsequent evolution. These impediments to abiogenesis, however, will be less critical for high mass K-type stars.

Quiescent M-type stars will have very low stationary concentrations of the fundamental molecules on their planets, $\sim 10^{-7}$ that of G-type stars for nominal $k=10^{-6}$ s$^{-1}$ (Table \ref{tab:ConcentrationTimes} and Figure \ref{fig:Stat_con_time}). Even the most highly active M-type stars will have stationary concentrations of the fundamental molecules less than $10^{-6}$ that of G stars. A four orders of magnitude lower photon intensity in the HCN production region $H$ of the spectrum compared to G stars (Table \ref{tab:Photon_fluxes_top_atm}) would mean a further decrease in the fundamental molecule stationary state concentrations, perhaps to only  $10^{-10}$ or $10^{-11}$ that of G-type stars. 

M-type and low mass K stars require very long times (months to years) to reach even these very low concentrations, implying high sensitivity to chemical degradation and vulnerability to extreme events. Frequent outbursts of high intensity X-rays and high energy particles could degrade precursors and the dissipatively structured molecules, as well as the planetary atmospheres. For example, active young M dwarfs produce high O$_2$ environments by dissociating H$_2$O, CO$_2$ or CO \cite{LugerBarnes2015}), which, besides drying the planet \cite{AirapetianEtAl2017}, inhibits the TDTOL mechanisms for abiogenesis and evolution by extinguishing soft UV-C light. 

High molar concentrations of the fundamental molecules, requires successive molecular transformations to occur only in the forward direction, i.e., towards product concentration profiles which are increasingly dissipative at each step \cite{Michaelian2026b}. This could not occur if the surface spectrum was rapidly time varying, as in the case of flaring stars. Furthermore, randomly timed transient intense soft UV-C fluxes are also unlikely to procure homochirality through the diurnal mechanism of UV-C and temperature assisted replication (UVTAR) of DNA/RNA under TDTOL described in reference \cite{Michaelian2018}.

Planets in the habitable zones of low mass stars could also become rapidly tidally locked, eliminating further evolution under the TDTOL which employs diurnal mechanisms, like UV-C and temperature assisted replication \cite{MichaelianSantillan2019} and the procurement of homochirality \cite{Michaelian2018}. 

Our analysis under the TDTOL, assumed Earth-like analogues with a secondary atmosphere similar to the best determinations of Earth's early Archean. Section \ref{sec:M-Flare} demonstrated that stationary molecular concentrations on planets of low mass stars are determined mainly by the intensity of light in the soft UV-C region and the chemical degradation rate constants, i.e., independently of atmospheric shielding of the hard UV-C region. However, for high mass stars, the effect of a chemically very different, thicker or thinner, atmosphere on the relevant photon fluxes at the planetary surface would have to be modeled and values determined for $P$ and $D$ (Table \ref{tab:Photon_fluxes_Earth_surf}) for each specific star-planet atmosphere combination. 

The greater the number of photochemical steps $n$ required in the production of a given fundamental biomolecule, the greater the dependency of the stationary concentrations on both the ratios of the photon fluxes $P/D$ and the cross sections for transformation $\alpha/\beta$, as seen by the dependence of the stationary state concentrations on these (Eq. (\ref{eq:Minfin})),
$$[\text{M}]_{\infty} \approx [\text{M}_0]\left(\frac{0.25\alpha P}{0.25\alpha P + 0.25\beta D+ k}\right)^{n-1} \frac{0.25\alpha P}{(0.25\beta D+k)}.$$
For example, for the nominal value of $k=10^{-6}$ s$^{-1}$, in going from $n=6$ to $n=8$ the ratio of stationary concentrations of the fundamental molecules for G-type stars over that for M-type stars goes from $10^7$ to $10^9$. Therefore, the more photons required in the dissipative structuring of a given molecule, the less likely it will be found on lower mass M or K-type stars compared to higher mass stars.   

Our analysis employed blackbody spectra, which tend to overestimate the true intensity in the UV-C region due to the presence of numerous absorption lines and molecular bands in real stellar atmospheres. These opacity sources, arising from ionized atoms and molecules such as CO, CN, OH, TiO, and Fe II, vary significantly in strength and wavelength bandwidth across different stellar types. In particular, lower mass cooler stars exhibit much stronger molecular absorption bands compared to G-type stars, leading to even greater suppression of the UV-C continuum in M dwarfs.

\section{Conclusion}

The Thermodynamic Dissipation Theory of the Origin of Life (TDTOL) asserts that life emerged on Earth through the non-equilibrium process of photochemical dissipative structuring of organic pigments in the soft UV-C region. Subsequent associations of these, and coupling with abiotic dissipative processes in the biosphere, was promoted by the thermodynamic imperative of increasing the dissipation of the full solar spectrum. From this perspective, we evaluated the suitability of different stellar spectral types for an Earth-like dissipative structuring abiogenesis and evolution. We considered Earth-analogues orbiting at the distance required to receive a top-of-atmosphere bolometric flux equal to the present solar constant, thereby ensuring comparable conditions for liquid water and photochemical reaction rates.

Under the TDTOL, and given the known physical properties of main-sequence stars, we conclude that the most promising targets for detecting microbial and thermodynamic biosignatures are Earth-like planets orbiting F-, G-, and higher-mass K-type stars. Biosignatures of complex or technological life are realistically expected only around G- and high-mass K-type stars. For M dwarfs and low mass K dwarfs, the combination of extremely low stationary state concentrations of fundamental biomolecules and long accumulation times (even for highly active stars), low precursor (e.g., HCN) concentrations, frequent high-energy atmosphere degrading flares, and tidal locking, renders a dissipative structuring origin and evolution of Earth-like life highly improbable. Only lithopanspermia from a F, G or K-type companion, after the onset of oxygenic photosynthesis, could plausibly seed such M worlds with life (e.g., in F/G/K+M binary systems).

We propose consideration of two complementary (to microbial) thermodynamic biosignatures rooted in TDTOL:
\begin{enumerate}
\item The global planetary entropy production rate derived from measurement of the planetary full incident and emitted spectra \cite{Michaelian2012a, MichaelianCano2022}.
\item A more readily observable proxy; an anomalously low soft-UV-C (205--285\,nm) albedo after correction for Rayleigh scattering and potential abiotic oxygen production.
\end{enumerate}

The second signature circumvents the severe contrast challenge posed by the high luminosity of F- and G-type hosts and directly probes the presence of UV-C-absorbing surface pigments, whether attributed to Archean-type organic chromophores or ozone derived from oxygenic photosynthesis, both of which TDTOL interprets as the result of molecular dissipative structuring.

The fact that life appears to be scarce within our solar system, and our galaxy, argues against the suggestion that a photochemical feedstock scenario is sufficient to bootstrap life, and instead argues in favor of the premise that abiogenesis is a delicate non-equilibrium thermodynamic dissipative structuring process where environmental conditions must be rather particular and stable, and probably similar to those of early Archean Earth.

We hope that this analysis encourages the astrobiology community to transcend feedstock scenarios (photochemical, chemical, or otherwise) and instead consider abiogenesis and evolution from the non-equilibrium photochemical dissipative structuring perspective founded on physical-chemical law. This could then help the community prioritize target selection for current and future searches for life, as well as guide the development of instruments sensitive to the detection of the fundamental thermodynamic imperative of life: the dissipation of stellar photons into heat.

\funding{This research was funded by DGAPA-UNAM, grant number IN104920.} 

\abbreviations{The following abbreviations are used in this manuscript:\\
\noindent 
\begin{tabular}{@{}ll}
AICN & 4-amino-1H-imidazole-5-carbonitrile\\
CO$_2$ & carbon dioxide\\
DNA    & deoxyribonucleic acid\\
FC & Franck-Condon region\\
FUV &  Far Ultraviolet - light within the region 100-200 nm\\
hard UV-C & light in the region 100-205 nm\\
HCN & hydrogen cyanide\\
H$_2$S & hydrogen sulfide\\
JWST & James Webb Space Telescope\\
NCCN & cyanogen\\
RNA  & ribonucleic acid\\
soft UV-C & light within the region 205-285 nm\\
TDTOL & Thermodynamic Dissipation Theory of the Origin of Life\\
UV-A  & light within the region 315-400 nm\\
UV-B  & light within the region 280-315 nm \\
UV-C  & light within the region 100-280 nm \\   
UVTAR & Ultraviolet and Temperature Assisted Replication - early enzyme-less replication of RNA and DNA 
\end{tabular}}

\reftitle{References}


\externalbibliography{yes}
\bibliography{Collection2026-2}

\end{document}